\documentclass[aps,twocolumn,groupedaddress,superscriptaddress,amsmath,amssymb,pra]{revtex4-2}
\usepackage{amsmath}
\usepackage{graphicx}
\usepackage{dcolumn}
\usepackage{bm}
\usepackage{bbold}
\usepackage[usenames,dvipsnames]{color}
\usepackage[colorlinks,citecolor=Blue,linkcolor=Red, urlcolor=Blue]{hyperref}
\usepackage{tikz}
\usetikzlibrary{decorations.pathmorphing}
\usetikzlibrary{arrows.meta}
\usepackage{braket}
\usepackage[normalem]{ulem}
\usepackage{tikz}

\begin{document}

\title{Counteracting dephasing in Molecular Nanomagnets by optimised qudit encodings}

\author{F. Petiziol}

\affiliation{Universit\`a di Parma, Dipartimento di Scienze Matematiche, Fisiche e Informatiche, I-43124 Parma, Italy}    
\affiliation{UdR Parma, INSTM, I-43124 Parma, Italy}

\author{A. Chiesa}
\affiliation{Universit\`a di Parma, Dipartimento di Scienze Matematiche, Fisiche e Informatiche, I-43124 Parma, Italy}    
\affiliation{UdR Parma, INSTM, I-43124 Parma, Italy}

\author{S. Wimberger}
\affiliation{Universit\`a di Parma, Dipartimento di Scienze Matematiche, Fisiche e Informatiche, I-43124 Parma, Italy}      
\affiliation{INFN, Sezione di Milano Bicocca, Gruppo Collegato di Parma, I-43124 Parma, Italy}

\author{P. Santini}
\affiliation{Universit\`a di Parma, Dipartimento di Scienze Matematiche, Fisiche e Informatiche, I-43124 Parma, Italy}      
\affiliation{UdR Parma, INSTM, I-43124 Parma, Italy}

\author{S. Carretta}
\email{stefano.carretta@unipr.it}
\affiliation{Universit\`a di Parma, Dipartimento di Scienze Matematiche, Fisiche e Informatiche, I-43124 Parma, Italy}     
\affiliation{UdR Parma, INSTM, I-43124 Parma, Italy}


\begin{abstract}
Molecular Nanomagnets may enable the implementation of qudit-based quantum error-correction codes which exploit the many spin levels naturally embedded in a single molecule, a promising step towards scalable quantum processors. 
To fully realize the potential of this approach, a microscopic understanding of the errors corrupting the quantum information encoded in a molecular qudit is essential, together with the development of tailor-made quantum error correction strategies.
We address these central points by first studying dephasing effects on the molecular spin qudit produced by the interaction with surrounding nuclear spins, which are the dominant source of errors at low temperatures.
Numerical quantum error correction codes are then constructed, by means of a systematic optimisation procedure based on simulations of the coupled system-bath dynamics, that provide a striking enhancement of the coherence time of the molecular computational unit. The sequence of pulses needed for the experimental implementation of the codes is finally proposed.
\end{abstract}                              
\maketitle

 \noindent{\bf INTRODUCTION}  \vspace{0.1cm} \\
Reliable quantum computation demands the adoption of strategies to protect the information being processed from external noise, \emph{i.e.}, of quantum error correction (QEC) schemes \cite{Terhal2015}. At the same time, while the ultimate quantum computer is expected to host QEC routines based on abstract, system-independent error models, the modern pioneering era of noisy intermediate-scale quantum devices \cite{Preskill2018, Tacchino2020} calls for strategies tailored for the specific physical hardware utilised. 

In their essence, QEC algorithms achieve information protection by suitably encoding an elementary two-state computational unit, a logical qubit, into a larger Hilbert space. This permits one to distinguish, and thus detect, different error occurrences without corrupting the information, so that it is then possible to retrieve it \cite{NielsenChuang2000}. 

While traditional QEC approaches realize the extra space by exploiting registers of many physical two-level systems, alternative routes to QEC have emerged wherein a logical qubit is hosted by a single many-level system (\emph{multi-level} or \emph{qudit} encodings) \cite{Gottesman2001,Troiani2012,Leghtas2013,Mirrahimi2014,Vlastakis2013,Linshu2017,Hu2019,Michael2016,Chiesa2020, Pirandola2008,Cafaro2012,Hussain2018}. 
The first advantage of the latter strategy is to prevent an overhead of physical units necessary to implement the code. Also, the manipulation of single or of pairs of logical qubits can be much easier, since they do not require controlling multiple physical objects at once \cite{Chiesa2021}. Moreover, the same multi-level object can provide protection against different types of errors \cite{Cafaro2012,Michael2016}. 

 A very promising architecture for the implementation of multi-level encodings is given by molecular nanomagnets \cite{Grover,Hussain2018,Chiesa2020}. Indeed, these highly coherent systems \cite{Bader2014,Zadrozny,Hill,Atzori2016,Atzori2017,Atzori2018,Atzori_JACS,Freedman_JACS,Freedman2014} offer many accessible spin levels \cite{Freedman_Ni,Freedman_Cr,Mn19powell,giant}, which can be manipulated with high accuracy through electromagnetic pulses, and they can be chemically engineered to meet desired requirements \cite{PRLLuis2011,PRLWedge,Aromi2012,Aguila2014,SciRepNi,Ardavan2015,modules,Chem,SIMqubit,VO2,Sessoli2019,Gaitarev,ErCeEr}. 

 The most important error in molecular spin systems at low temperature is given by pure dephasing, that is, the suppression of coherence between different spin states. 
Such a decoherence mechanism originates principally from the hyperfine interaction of the central (electronic or nuclear) molecular spin with the bath of surrounding nuclear spins \cite{Troiani2008,Ghirri2015,Chen2020}. Except from specific situations \cite{Coish2008}, decoherence of a central spin induced by a nuclear spin bath is known to produce non-exponential decay behaviour \cite{Klauder1962,Abe2004,Witzel2005,Ardavan2007,Troiani2008,Bader2014,Graham2017,Chen2020}. This is due to many factors, such as non-negligible entanglement building up between the spin and an evolving bath, the limited number of nuclear spins ($\sim 10^2$) usually surrounding the molecular spin $\mathcal{S}$, the slow relaxation timescales of the bath relative to the motion of the central spin. 
Although mandatory to design targeted codes, a QEC framework which takes into account both the multi-level nature of a spin $S$ larger than 1/2 and the explicit structure and dynamics of the nuclear spin bath is still missing. 

 In this work, we develop a class of numerical spin qudit codes which are designed based on a the detailed microscopic structure of the environment responsible for errors {and which provide a strongly enhanced correction efficiency}. Moreover, the sequence of control pulses and measurements needed for an experimental implementation of such codes is discussed. While the advantage of these codes as compared to a simple spin 1/2 is evident already using small $S$, the performance strikingly improves as qudits with larger spin are used, thus positively exploiting more and more available levels in the molecular spectrum.  
 
 The codes are derived by first analysing the decoherence effects experienced by a qudit spin $S>1/2$ embedded into a realiztic nuclear spin bath, by means of numerical simulations of the coupled qudit-bath dynamics through a cluster-correlation expansion \cite{Yang2008}. %
A systematic procedure is then put forward to capture the spin-dephasing process by means of error operators acting on the system, which are then used to derive optimised code words for QEC. 
Thanks to the flexibility of the procedure, the numerical codes can be optimised taking into account the specific timescale of free evolution admitted between two subsequent QEC cycles, thus allowing one to largely reduce the number of correction steps sufficient to ensure a desired fidelity. As such, they are an optimal candidate for realizations in near-term devices, in which the implementation of the QEC can be noisy and can take relatively long times. \\

\begin{figure}[b]
\includegraphics[width=0.8\linewidth]{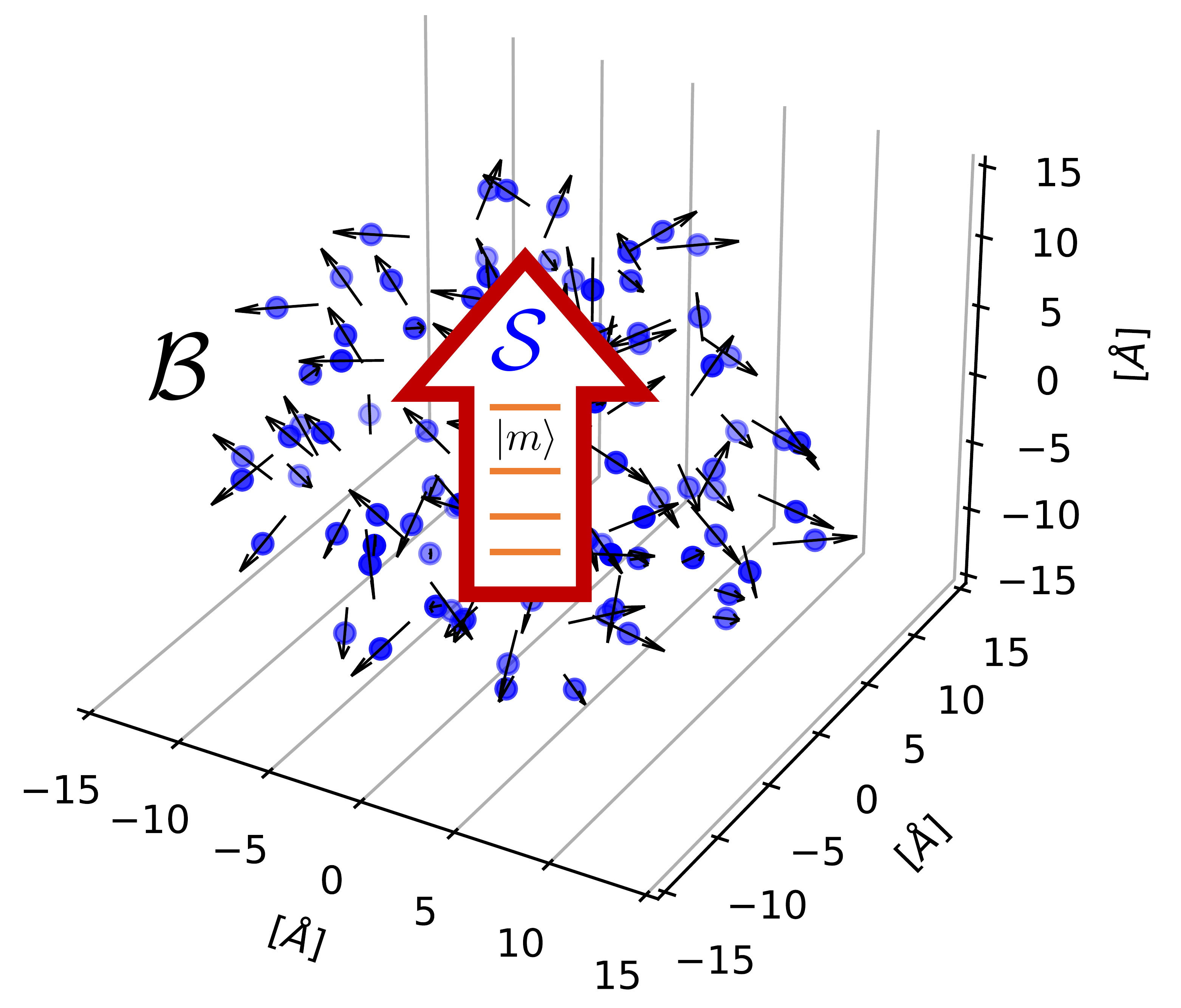}
\caption{ \textbf{Model system}. A spin $\mathcal{S}$ larger than 1/2, whose many-level structure is exploited for performing multi-level encodings, interacts with the bath $\mathcal{B}$ of surrounding nuclear spins. Entanglement between $\mathcal{S}$ and $\mathcal{B}$ induces spin dephasing, which is counteracted through quantum error correction. Nuclear spins are plotted from a sample configuration of 100 nuclear spins used in the simulations, generated randomly within a sphere of radius $15 \, \AA$ and with a minimal distance of $3 \,\AA$ between spins (see `Methods').}
\label{fig0}
\end{figure}

\noindent {\bf RESULTS} \vspace{0.1cm}

\noindent \textbf{Physical system and decoherence mechanisms} \\%
The system analyzed in the following is a molecular electronic spin $\mathcal{S}$ (sketched in Fig. \ref{fig0}), described by the Hamiltonian $\hat{H}_{\mathcal{S}}= D \hat{S}_z^2 + \Omega \hat{S}_z$. Here, $\{\hat{S}_x, \hat{S}_y, \hat{S}_z\}$ are spin $S>1$ operators, with eigenstates of $\hat{S}_z$ being defined by $\hat{S}_z\ket{m} = m \ket{m}$. The parameter $D$ indicates the zero-field splitting (assumed to be axial, for simplicity) and $\Omega = g_z \mu_\mathrm{B} B_z$, with $g_z$ the longitudinal $g$-factor and $\mu_\mathrm{B}$ the Bohr magneton, characterises the Zeeman interaction with a static magnetic field along the $z$ direction. 
The analysis developed here can apply both to a qudit given by a single spin-$S$ ion and to a giant spin originating from the strong exchange interactions between different ions \cite{Chiesa2017}. Also, while we focus here on the case of an electronic qudit, the same treatment can also be applied to a nuclear qudit with small adaptations commented in `Methods'.

 For a molecular electronic spin $\mathcal{S}$ at low temperature, the hyperfine coupling with the surrounding nuclear spin bath is the dominant source of decoherence. Indeed, as typically done in quantum computing platforms, we assume to work at temperatures much smaller than the relevant energy scales of the qudit ($\Omega,D k_\mathrm{B}^{-1} \sim$ K), such that thermal population of the excited states is negligible. In these conditions, phonon absorption is very unlikely. At the same time, the qudit energy gaps are much smaller than the Debye energy (typically in the $\gtrsim 50$ K range), thus making also resonant phonon emission (whose probability scales as the third power of the gap) negligible \cite{Wurger}. 
 In general, the effect of spin-phonon coupling on the system dynamics can be calculated from diagonalisation of the rate matrix \cite{Wurger}, yielding a decay of both diagonal and off-diagonal elements of the system density matrix (associated to relaxation and dephasing, respectively) on related time-scales. In particular, phonon-induced dephasing is limited by the relaxation time $T_1$. A body of experiments on molecular spin qubits and qudits \cite{SIMqubit,BaderChemComm,Freedman_Cr} demonstrate that this is not the case at low temperature, where $T_1$ increases exponentially and becomes several orders of magnitude longer than the dephasing time. Hence, at low temperature phonons are not responsible for pure dephasing on the spin system and other mechanisms come into play. 

Spin dipole-dipole interactions between electronic spins can have an important effect in concentrated samples, but this is strongly reduced if magnetic centres are diluted in a diamagnetic matrix~\cite{Hill} and is not relevant here, because we consider a perspective architecture consisting of a single (or a few) molecular unit(s) \cite{Takahashi,Ghirri2015}. 
 
We therefore focus on the bath $\mathcal{B}$ of nuclear spins surrounding $\mathcal{S}$. The number of nuclear spins in the bath may range from a few tens to a few hundreds in realiztic molecular complexes, thus being rather far from the infinite-bath limit underpinning typical Markovianity approximations. 
By working in the so-called `coherence window' \cite{Stamp}, in which the system energy gaps are much larger than the gaps of the nuclear spin bath, off-diagonal operators inducing population transfer on the system can be neglected. The system-bath dynamics can be described in this regime by effective spin Hamiltonians featuring only a diagonal coupling between $\mathcal{S}$ and the bath, which are derived via perturbation theory. This type of Hamiltonians has been studied in the context of a (pseudo)spin $S=1/2$ interacting with a nuclear spin bath \cite{Yao2006,Troiani2008, Troiani2012}. 
In `Methods', we deduce an effective Hamiltonian for the dynamics of a generic spin $S>1/2$. In interaction picture with respect to $\hat{H}_\mathcal{S}$ and to first order in $\Omega^{-1}$, this Hamiltonian is of the form
\begin{equation}\label{eq:ham_eff}
\hat{H} = \hat{H}_{\mathcal{B}}^{(0)} + \hat{S}_z \hat{H}_{\mathcal{B}}^{(1)} + (\hat{S}^2 - \hat{S}_z^2) \hat{H}_{\mathcal{B}}^{(2)}.
\end{equation}
Both the intrinsic and the qudit-conditioned Hamiltonians $H_{\mathcal{B}}^{(k)}$ of the bath can be written in the general form
\begin{align}\label{eq:ham_det}
H_{\mathcal{B}}^{(k)} = & \sum_{n=1}^{N} \left(a_n^{(k)}  \hat{I}_n^z + b_{n}^{(k)} (\hat{I}_n^z)^2\right)  \nonumber \\
&+ \sum_{ \substack{n,m=1\\m\ne n}}^N  \left( c_{n,m}^{(k)} \hat{I}_n^+ \hat{I}_m^-  + d^{(k)}_{n,m} \hat{I}_n^z \hat{I}_m^z \right),
\end{align}
where $ \{\hat{I}_k^x, \hat{I}_k^y, \hat{I}_k^z\}$ are spin operators for the $k$-th nuclear spin of the bath and $\hat{I}_k^{\pm} = \hat{I}_k^x \pm \mathrm{i} \hat{I}_k^y$. The coefficients $a_{n,m}^{(k)}, b_{n,m}^{(k)}, c_{n,m}^{(k)}, d_{n,m}^{(k)}$ are a function of the hyperfine couplings between $\mathcal{S}$ and $\mathcal{B}$, the nuclear spin-spin dipolar couplings, and $\Omega$. In the following, nuclear spins are assumed to be protons ($I=1/2$), since hydrogen nuclei typically represent the major decoherence source. Other relevant parameters and details on the simulations are given in `Methods'. 

 System-bath entanglement, generated by the Hamiltonian of Eq. \eqref{eq:ham_eff}, can be interpreted in terms of `which-way information' accumulated in the state of the nuclear spins: depending on the state $\ket{m}$ of $\mathcal{S}$, the bath undergoes different interacting evolutions described by Hamiltonians $\hat{H}_{\mathcal{B},m} = \bra{m} \hat{H} \ket{m}$.
These conditioned bath evolutions result in a decay of coherences in the system density matrix $\rho_{\mathcal{S}}(t)$, according to $\bra{n} \rho_{\mathcal{S}}(t) \ket{m} = L_{nm}(t) \bra{n} \rho_{\mathcal{S}} (0) \ket{m}$. The function $L_{nm}(t)=\text{tr}_\mathcal{B} \left[ \mathrm{e}^{- \mathrm{i} \hat{H}_{\mathcal{B},n} t} \rho_\mathcal{B}(0) \mathrm{e}^{\mathrm{i} \hat{H}_{\mathcal{B},m} t} \right]$, with $\rho_\mathcal{B}(0)$ the initial bath state, is computed numerically in the following through a cluster-correlation expansion (CCE) \cite{Yang2008,Yang2009}. 

\begin{figure}[t]
\includegraphics[width=\linewidth]{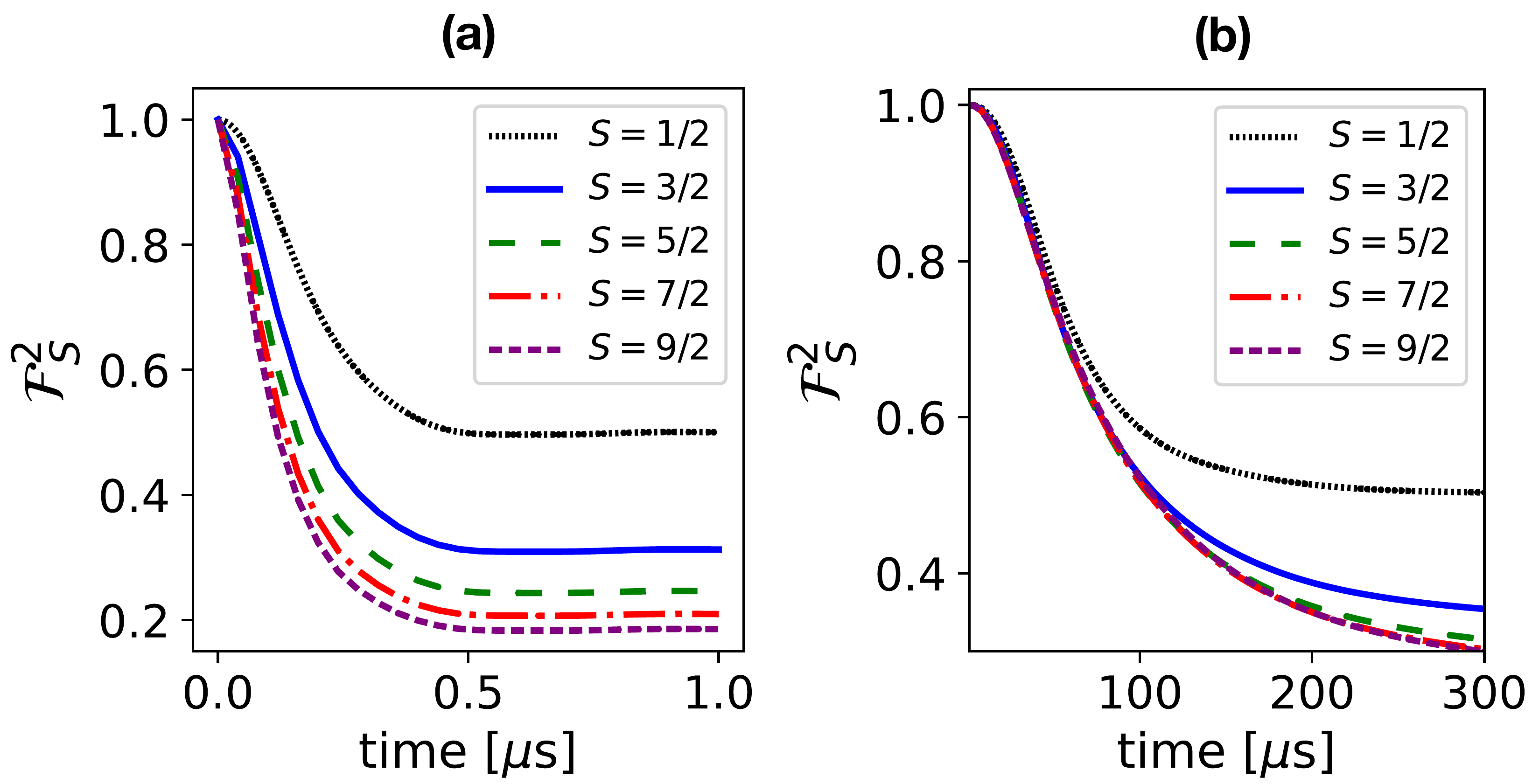}
\caption{ {\bf Inhomogeneous broadening and spin echo. (a)} Decay of the squared fidelity with respect to the initial state under inhomogeneous broadening for different spins $S$ initialised in a state $\ket{\psi_L} = (\ket{0_L}+ \mathrm{i} \ket{1_L})/\sqrt(2)$, where $\ket{0_L}$ and $\ket{1_L}$ are spin-binomial code words  corresponding to spin $S$ \cite{Chiesa2020} (see also `Methods'); the results have been averaged over $2^6$ initial spatial configurations of the nuclear spins, as explained in `Methods'. {\bf(b)} Decay of the echo squared fidelity for the same initialisation of (a). For each time point $t$, a generalised echo transformation of the form $\mathrm{e}^{-\mathrm{i} \pi \hat{S}_x}$ at $t/2$ is understood.}
\label{fig:IB_echo}
\end{figure}

 In a free-decay experiment, the main decoherence process is given by inhomogeneous broadening \cite{Klauder1962}. The system-bath diagonal coupling $a_n^{(1)} \hat{S}_z \hat{I}_k^z$ has the effect of a classical random magnetic field --- the Overhauser field --- on $\mathcal{S}$. Uncertainty in the actual bath state then produces, for the density matrix $\rho_{\mathcal{S}}(t)$ of the qudit, a Gaussian decay for the single transition coherence, 
\begin{equation}
\bra{m} \rho_{\mathcal{S}}(t) \ket{n} \approx \mathrm{e}^{-(n-m)^2 \Gamma^2 t^2 } \bra{m} \rho_{\mathcal{S}}(0)\ket{n},
\end{equation}
with $\Gamma^2 = \sum_k (a_n^{(1)})^2/4$ (see `Methods'), over timescales much faster than those set by the nuclear spin-spin interaction. This is shown in Fig. \ref{fig:IB_echo}, where the squared fidelity $\mathcal{F}^2_S(t)$ with respect to the initial state, with $\mathcal{F}_S(t) =\text{tr}_\mathcal{S} \sqrt{\sqrt{\rho_\mathcal{S}(t)} \rho_\mathcal{S}(0) \sqrt{\rho_\mathcal{S}(t)}}$ \cite{NielsenChuang2000}, is depicted for different spin $S$. The fidelity decays over timescales of hundreds of nanoseconds. 

 The dramatic effect of inhomogeneous broadening on the spin coherence is routinely compensated for in experiments by means of spin echo/refocusing schemes, whose basic example is the Hahn echo. For different qudit spins, the echo dynamics is shown in Fig. \ref{fig:IB_echo}(b). The realization of the echo transformations is further described in `Practical Implementation'. The coherence decays over timescales of tens-to-hundreds of microseconds, signalling that the effect of inhomogeneous broadening is removed to large extent. The decay is now due to the quantum dynamics of the bath, and is mainly determined by the contribution given by intra-bath interactions in $H_{\mathcal{B}}^{(0)}$, of the form $c_{n,m}^{(0)} \hat{I}_n^+ \hat{I}_m^-$. If the latter were absent, the echo would recover unit fidelity independently from $S$ over timescales of hundreds of microseconds, until virtual flip-flops described by terms of type $c_{n,m}^{(1)} \hat{S}_z\hat{I}_n^+ \hat{I}_m^{-}$ set in. This effect is still partially visible in Fig. \ref{fig:IB_echo}(b) in the fact that, for short timescales with respect to interactions in $H_{\mathcal{B}}^{(0)}$, the fidelity exhibits almost overlapping decay for different $S$.\\ 

\noindent\textbf{Optimised qudit encoding}

\noindent While increasingly sophisticated echo pulse sequences can recover the effect of inhomogeneous broadening to a better and better degree, the spin coherence remains irremediably affected by the interacting quantum dynamics of the bath. We derive in the following qudit QEC codes as a means to protect the system from these effects.
In particular, we develop a framework for designing optimal numerical codes which are based on the detailed description of the system-bath dynamics adopted in this work.

 A quantum error correction code can be defined by following two fundamental steps. The first step is to identify error operators $\hat{E}_k$ which describe the effect of the noise source on the system, \emph{i.e.}, such that the state of $\mathcal{S}$ at time $t$ can be related to the initial state through
\begin{equation}\label{eq:err_ops}
\hat{\rho}_{\mathcal{S}}(t) = \sum_k \hat{E}_k \hat{\rho}_{\mathcal{S}}(0) \hat{E}_k^\dagger.
\end{equation}
The second step is the derivation of computational states $\ket{0_L}$ and $\ket{1_L}$ that satisfy Knill-Laflamme conditions for quantum error correction \cite{Knill1997}, namely, for all $k$ and $j$,

\begin{eqnarray}
& \bra{0_L} \hat{E}_k^\dagger \hat{E}_j \ket{0_L} = \bra{1_L} \hat{E}_k^\dagger \hat{E}_j \ket{1_L}, \label{eq:kl}\\
& \bra{0_L} \hat{E}_k^\dagger \hat{E}_j \ket{1_L} = 0. \label{eq:kl2}
\end{eqnarray}

These conditions demand that, when errors $\hat{E}_k$ affect an initial state $\ket{\psi_L}=\alpha \ket{0_L} + \beta \ket{1_L}$, the code words are modified but the corresponding coefficients $\alpha$ and $\beta$ are not, and the information they carry is thus preserved. Moreover, errors do not create ambiguity between $\ket{0_L}$ and $\ket{1_L}$: error words $\hat{E}_k \ket{w_L}$ span subspaces that are orthogonal to each other for different $w=0,1$.\\
 If these conditions are fulfilled, then different errors $\hat{E}_k$ can be distinguished, detected and corrected. In practice, a measurement is devised whose outcome discriminates the error occurrence and a recovery operation restores the initial state. Importantly, the $2S+1$ levels of a spin $S$ offer enough state space to detect and correct a number $N_\textrm{corr} = \lfloor S \rfloor$ of error operators $\hat{E}_k$ \cite{Chiesa2020}, where $\lfloor S \rfloor$ indicates the largest integer smaller or equal than $S$.
Given that a decomposition of the form \eqref{eq:err_ops} involves in general a larger number of $\hat{E}_k$, it is essential to identify the $N_\textrm{corr}$ error operators which have stronger effect, such that the code can be tailored for them ensuring optimal correction. 
\begin{figure}[t]
\includegraphics[width=\linewidth]{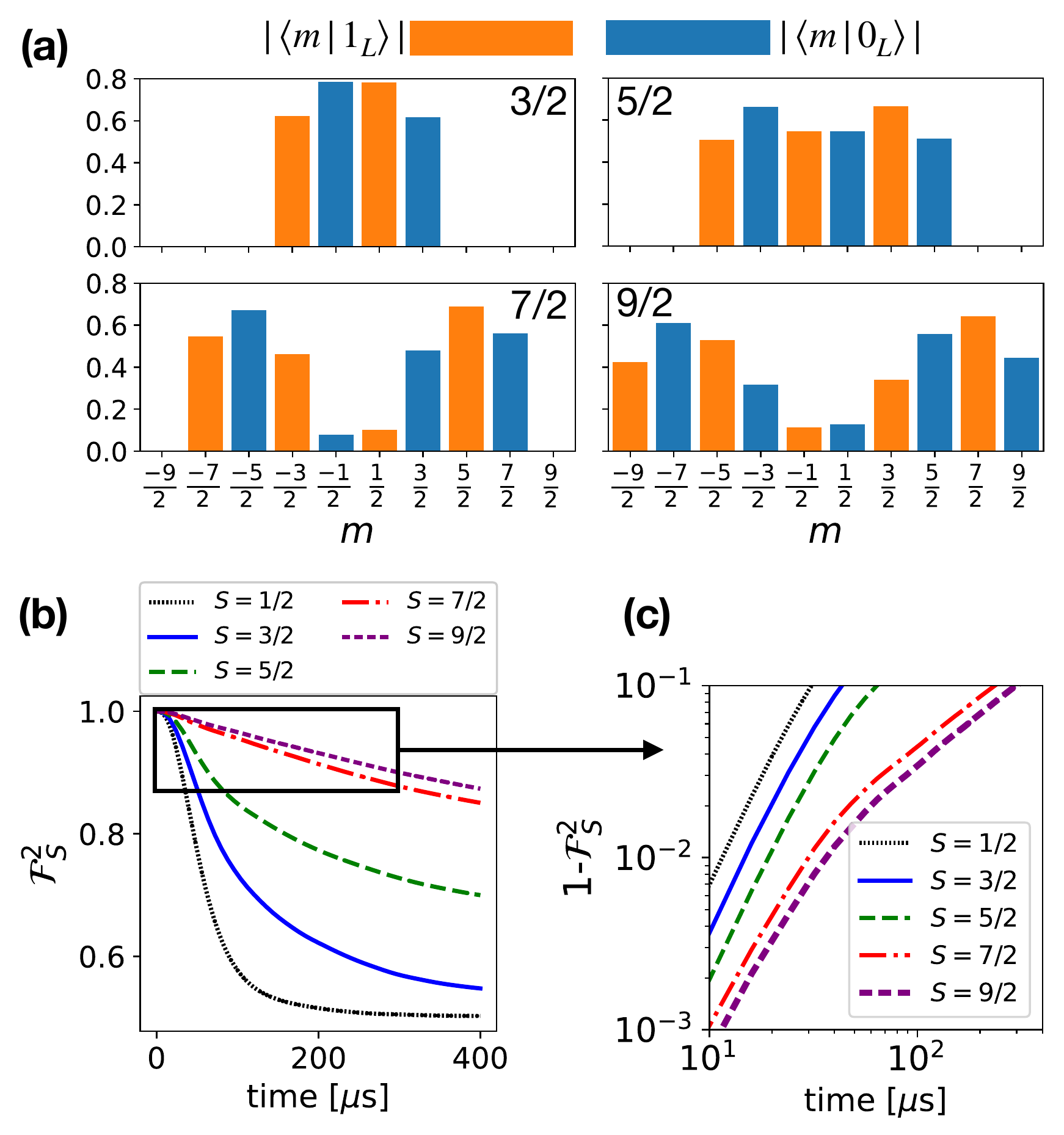}
\caption{ \textsc{Figure 3.} {\bf Numerically optimised qudit codes}. \textbf{(a)} Absolute value of the overlap of the optimised code-words $\ket{0_L}$ (blue) and $\ket{1_L}$ (orange) with each state $\ket{m}$ for different spins $S$. \textbf{(b)} Squared average fidelity (over nuclear configurations) for different qudit spins $S$, for initial state $\ket{\psi_L} = [\ket{0_L} + \mathrm{i} \ket{1_L}]\sqrt{2}$ and numerical code words $\ket{0_L}$ and $\ket{1_L}$ optimised at $t=5~\mu$s. \textbf{(b)} Infidelity $1-\mathcal{F}_S^2$ for an inset of panel (b) representing the region of $\mathcal{F}_S^2\ge 0.9$.}
\label{fig:opt1}
\end{figure}

For the spin-dephasing scenario considered here, a decomposition of the form \eqref{eq:err_ops} with exact error operators is not known, thus preventing a derivation of adequate code words for this type of noise. In order to overcome this limitation, we introduce an iterative numerical optimisation procedure which, given $\hat{\rho}_\mathcal{S}(0)$ and $\hat{\rho}_\mathcal{S}(t)$ computed through CCE, aims at determining a number $N_\textrm{corr}$ of operators $\hat{E}_k$ by decreasing contribution to $\hat{\rho}_\mathcal{S}(t)$.
Starting from $\hat{\rho}_\mathcal{S}^{(0)} \equiv \hat{E}_0 \hat{\rho}_{\mathcal{S}}(0) \hat{E}_0^\dagger$, the $n$-step estimate $\hat{\rho}_{\mathcal{S}}^{(n)}$ to $\hat{\rho}_{\mathcal{S}}(t)$ of the iteration is defined according to
\begin{equation}\label{eq:iteration}
\hat{\rho}_{\mathcal{S}}^{(n)} = \hat{\rho}_\mathcal{S}^{(n-1)} + \hat{E}_n \hat{\rho}_\mathcal{S}(0) \hat{E}_n^\dagger. 
\end{equation}
At the $n$-th step, the distance $\lVert \rho_\mathcal{S}(t)-\rho_\mathcal{S}^{(n-1)} \lVert$ (here, $\Vert \cdot \Vert$ is the Hilbert-Schmidt norm) is numerically minimised in order to find optimal parameters for a parametrised form or $\hat{E}_n$ (specified in the following), and the outcome is used for the subsequent step of the iteration. If a hierarchy of $\hat{E}_k$ exists, a successful optimisation will return error operators which give less and less contribution, in norm, to the density matrix. In this sense, the numerical procedure then leads to an optimal decomposition of $\hat{\rho}_\mathcal{S}(t)$ in the form \eqref{eq:err_ops}.

 From the structure of the system-bath Hamiltonian $\hat{H}$ of Eq. \eqref{eq:ham_eff}, it follows that $\rho_{\mathcal{S}}(t)$ can be generically written in the form \eqref{eq:err_ops} with error operators that are diagonal in the basis of states $\ket{m}$. 
Given that the Hilbert space of $\mathcal{S}$ is finite with dimension $2S+1$, the error operators $\hat{E}_k$ can be expanded onto a basis $\{ \hat{D}_m \}$ of $2S+1$ diagonal operators. Indeed, a diagonal matrix in this space can have at most $2S+1$ linearly independent entries.
This justifies an expansion for the error operators of the form
\begin{equation}\label{eq:err_ansatz}
\hat{E}_k = \sum_{m=0}^{2S} E_{k,m} \hat{D}_m.
\end{equation}
\begin{figure}[t]
\includegraphics[width=\linewidth]{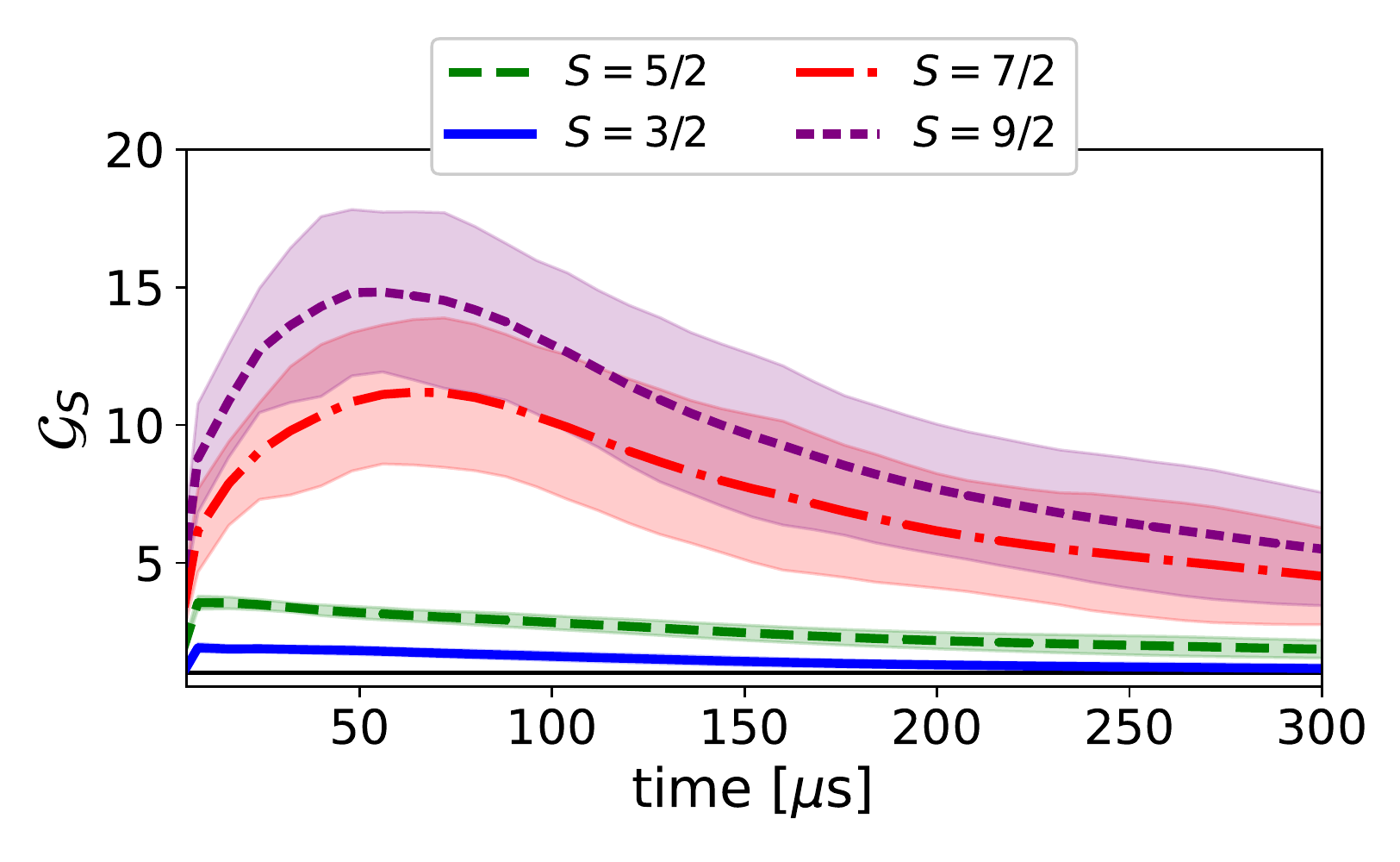}
\caption{\textbf{Gain}, defined in Eq. \eqref{eq:gain}, with respect to non-corrected spin 1/2 for the data in Fig. \ref{fig:opt1}. The curves indicate the average of $\mathcal{G}_S(t)$ over the different nuclear spin spatial configurations (generated as explained in `Methods'), while the shaded areas mark the region included between $\mathcal{G}_S(t) \pm \sigma_{S}(t)$, where $\sigma_S(t) = \sqrt{\langle \mathcal{G}_S^2(t)\rangle - \langle \mathcal{G}_S(t)\rangle^2}$ is the standard deviation and $\langle \cdot \rangle$ denotes averaging. The shaded areas show that a large gain is obtained for all spatial configurations at given spin $S$. The standard deviation $\sigma_S(t)$ increases with $S$ since the denominators in Eq. \eqref{eq:gain} become smaller and smaller for increasing $S$. A comparison with spin-binomial codes is further given in `Methods'.}
\label{fig:gainopt}
\end{figure}
The coefficients $E_{k,m}$ are the free parameters for the numerical optimisation.
The basis $\{ \hat{D}_m \}$ is chosen in the following to be given by the projectors $\hat{D}_m = \ket{m}\!\bra{m}$ over the $\ket{m}$ states.
Once the error operators are found, the code words enabling their quantum error correction are determined by imposing Knill-Laflamme's conditions \eqref{eq:kl} numerically, as detailed in `Methods'. The code words obtained from this procedure are depicted in Fig. \ref{fig:opt1}(a) for values of spin from $S=3/2$ to $S=9/2$. By construction, $\ket{0_L}$ and $\ket{1_L}$ have support on different subsets of $\ket{m}$ states in an alternate fashion. This automatically guarantees the fulfilment of Knill-Laflamme's condition \eqref{eq:kl2}, reducing the number of free parameters for the numerical search required to impose condition \eqref{eq:kl}.

 The performance of the optimised qudit codes is remarkable, as shown in Fig. \ref{fig:opt1}, where the fidelity after the QEC is reported, for a set of codewords corresponding to different qudit spin $S$. In particular, for each time $t$, Fig. \ref{fig:opt1}(b) represents the squared fidelity of the recovered state with respect to the encoded state, achieved by performing an instantaneous QEC at time $t$. In panel \ref{fig:opt1}(c), we report instead the infidelity $1-\mathcal{F}_S^2(t)$ in log-log scale for an inset of panel (b). One can observe that, while a squared fidelity above 0.9 is maintained for a spin 1/2 only up to $\sim 30~\mu s$, the recovered fidelity is above that value for as long as $\sim$40, 65, 240 and 300 $\mu$s for qudit spin $S=3/2$, 5/2, 7/2 and 9/2, respectively. Similarly, the same spin values guarantee a recovered fidelity above 0.99 for up to $\sim$15, 20, 29 and 37 $\mu$s, well longer than the spin 1/2, $\sim10$ $\mu$s. The possibility to recover high fidelity even after rather long evolution times is a crucial resource for near-term implementations. Indeed, if the rate at which subsequent QEC cycles need to be done is too large, the advantage of the correction may get lost in a realiztic implementation because of the non-negligible time necessary to implement all the measurements and recovery operations of the QEC step.
 
 The substantial advantage in increasing the spin of the qudit is further emphasised in Fig. \ref{fig:gainopt}, where the gain with respect to the spin 1/2,
\begin{equation}\label{eq:gain}
\mathcal{G}_S(t) = \frac{1-\mathcal{F}_{1/2}^2(t)}{1 - \mathcal{F}_S^2(t)},
\end{equation}
 is reported as a function of time, for different values of the spin $S$. A remarkable maximal gain, larger than $10$, is attained, \emph{e.g.}, for a spin $7/2$ at around $60$ $\mu$s, and a maximal gain around $15$ is attained for $S=9/2$. 
 
\begin{figure}[t]
\includegraphics[width=\linewidth]{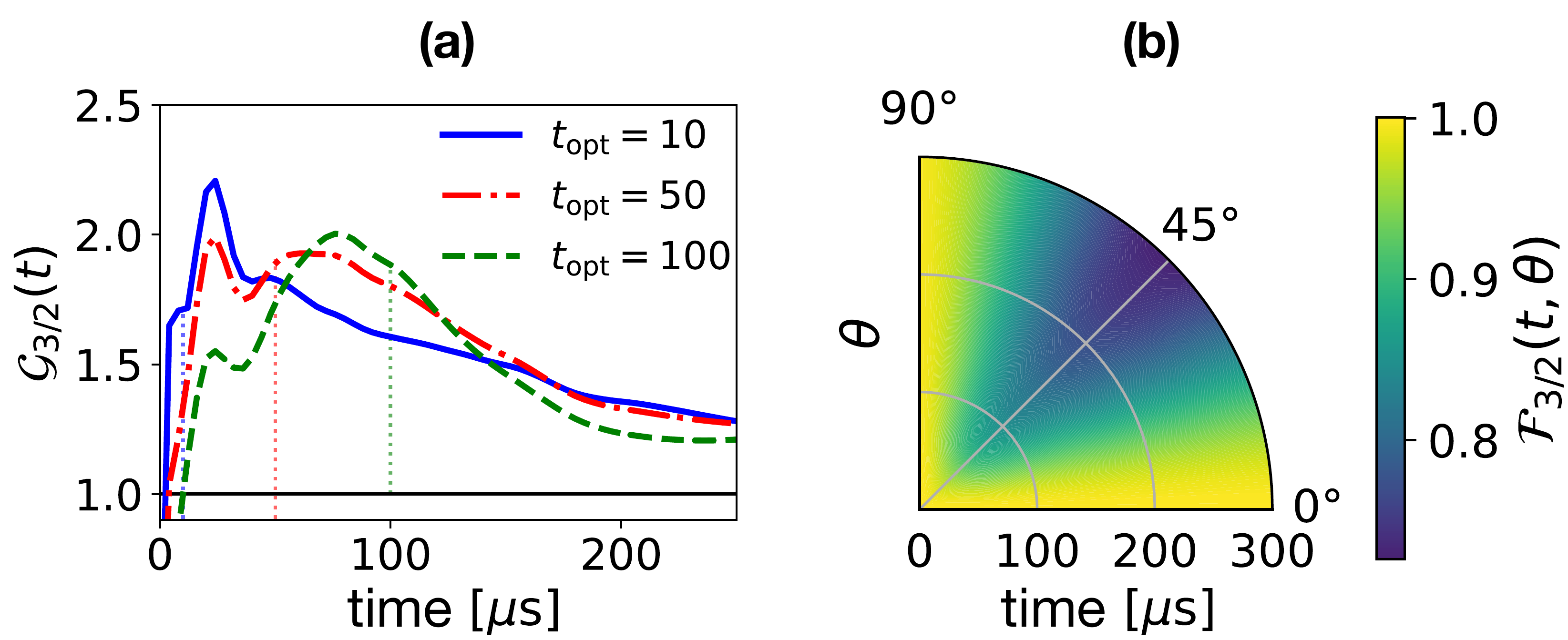}
\caption{{\bf Optimisation time and initial states. (a)} Gain as a function of time for a logical state $\ket{\psi_L}=(\ket{0_L} + \mathrm{i} \ket{1_L})/\sqrt{2}$ for numerical codewords optimised at three different times $t_\textrm{opt}= 10, 50, 100\, \mu$s and $S=3/2$. 
{\bf(b)} Fidelity (color scale) as a function of time (radial scale) for different initial superposition states $\cos\theta\ket{0_L} + \mathrm{i} \sin\theta\ket{1_L}$ ($\theta$ depicted in angular scale) of a set of optimised numerical code words for $S=3/2$. }
\label{fig:opt2}
\end{figure} 
 
Depending on the time at which the optimisation is performed, rather different numerical code words can be obtained, reflecting the interplay of different interaction scales in the system-bath Hamiltonian. However, broad temporal windows can be recognised, in which the code words maintain essentially the same structure, while being quite different in two different regimes. Therefore, a given set of code words maintains a stable performance if the QEC is implemented in a rather broad time interval around the optimisation time. These features can be observed in Fig. \ref{fig:opt2}(a), where the gain as a function of time is shown for three different code word pairs obtained by optimising at times $t_{\textrm{opt}}=10,50,100$ $\mu$s for $S=3/2$. As expected, while a unique pair giving largest gain at all times cannot be found, code words optimised at a given time provide very good performance in a rather large region around that time. 
We have finally checked that the performance of the numerically optimised code words does not critically depend on the initial state chosen for our procedure, as demonstrated in Fig. \ref{fig:opt2}(b) for the exemplary case of $S=3/2$.
The squared fidelity (color scale) as a function of time (radial scale) is reported for different values of the angle $\theta$ (angular scale) characterising an initial state of the form $\cos(\theta) \ket{0_L} + \mathrm{i} \sin(\theta) \ket{1_L}$. Large fidelities are attained for all initial states, with fidelity increasing as one departs from the state with equal weights [$\theta=\pi/4$], that is, the most decoherence-prone state, used in the other reported simulations. \\

 \noindent \textbf{Practical implementation}
 \begin{figure*}[t]
\includegraphics[width=0.8\linewidth]{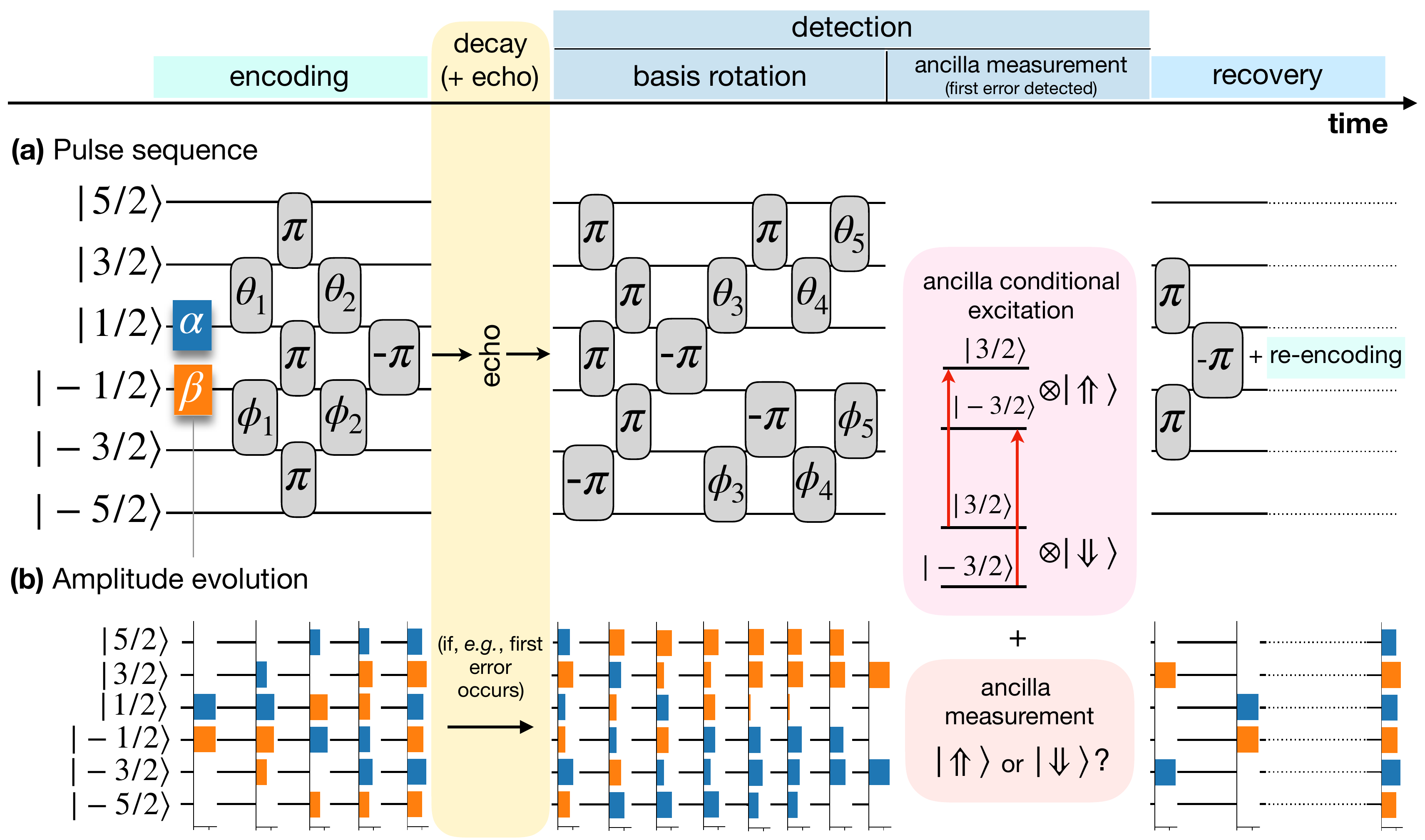}
\caption{{\bf Pulse sequence for $\bm{S=5/2}$}. {\bf (a)} Explicit sequence of pulses to implement the QEC for a spin $S=5/2$. Each horizontal line represents a spin state $\ket{m}$ with time flowing from left to right. Each grey box indicates a pulse implementing a rotation $Y_m(\theta)$ as described in the text, with the values of the angles $\theta_k$ and $\phi_k$ given in Table \ref{tab:1} in `Methods'. {\bf (b)} Time evolution of the absolute value of the amplitude $\lvert\braket{m \vert \psi(t)  }\!\lvert$ of the qudit state $\ket{\psi(t)}$ over each $\ket{m}$ state at different stages of the control sequence, corresponding to the control pulses depicted in (a). The sequence holds for any choice of $\alpha$ and $\beta$, though panel (b) shows an example with $\alpha=1/\sqrt{2}$, $\beta=\mathrm{i}/\sqrt{2}$. Blue and orange colours represent amplitudes associated to $\alpha$ and $\beta$ ({\it i.e.}, $\ket{0_L}$ and $\ket{1_L}$), respectively. To exemplify the effect of the basis rotation and the outcome of the measurement during the detection, the first-error case (related to the error operator $\hat{E}_1$ of Eq. \eqref{eq:err_ansatz}) is shown in (b). }
\label{fig:pulses}
\end{figure*} 

\noindent Having analysed the ideal efficiency of the optimised qudit codes, we now turn to a discussion of how to implement in practice all the steps of the quantum error correction procedure, namely encoding, detection and recovery, whose formal description is given in `Methods'.
The manipulation of a spin $S>1/2$ system requires of course more complex control sequences compared to a spin $S=1/2$. Nonetheless, it can be realized in a total time sufficiently short so that it does not significantly impact the efficiency of the ideal QEC, as detailed in the following.  
 
The population transfers required among spin states can be realized using sequences of resonant microwave/radiofrequency pulses. These are described in the Hamiltonian by time-dependent control fields of the form $g_y \mu_\mathrm{B} B_y(t) \cos(\omega t) \hat{S}_y$ where the envelope $B_y(t)$ is typically rectangular or Gaussian. These pulses induce transitions from a spin state $\ket{m}$ to $\ket{m\pm 1}$ when $\omega$ is set at the corresponding transition frequency, implementing a two-state unitary rotation $Y_m(\theta) = \exp[\theta/2(\ket{m+1}\bra{m} - \ket{m}\bra{m+1})]$ between states $\ket{m}$ and $\ket{m+1}$ of an arbitrary angle $\theta$.
All the steps of the QEC are illustrated in the following, and the explicit realization for a $S=5/2$ qudit code is depicted in Fig. \ref{fig:pulses}. The implementation proposed here generalises the one proposed for spin-binomial codes \cite{Chiesa2020}. \\

 {\it Encoding.} We assume that the information, {\it i.e.} the coefficients $\alpha$ and $\beta$, is initially encoded in a simple state such as $\alpha \ket{-1/2} + \beta\ket{1/2}$. The preparation of the logical state $\alpha \ket{0_L} + \beta \ket{1_L}$ for arbitrary $\alpha$ and $\beta$ is then realized by alternating pulses $Y_m(\theta)$ distributing population among the different $\ket{m}$ states and $\pi$-pulses $Y_m(\pm\pi)$ which rearrange the different populations in the correct order. The angles $\theta$ of the two-level rotations are fully determined by the components of the code-words on each $\ket{m}$ state. The explicit sequence for $S=5/2$ is given in panel `encoding' of Fig \ref{fig:pulses}(a) with the angles given in Table \ref{tab:1} of `Methods'. Once the state has been codified, the system is left to decay freely for a time $t/2$, then a spin echo pulse sequence is performed, and the quantum error correction finally takes place at time $t$ starting with the error detection. These pulse sequences can also be used to perform single-qubit gates between the code-words, for instance by first re-mapping the code-words to, e.g. $\ket{\pm1/2}$ states, performing the desired two-state operation, and re-encoding.

 {\it Spin echo.} The Hahn echo for a spin 1/2 can be understood as a magnetic pulse along $x$ or $y$ which effectively flips the spin. Then, the spin can be viewed as effectively evolving with $\hat{H}$ for a time $t/2$ and
 with the same Hamiltonian but with $\hat{S}_z$ changed to $-\hat{S_z}$ for an equal time $t/2$, where $t$ is the time at which the QEC is performed. Similarly, the echo scheme is extended here to a larger spin $S>1/2$ by considering a `generalised-pulse' transformation of the form $U_{\mathrm{echo}} = \mathrm{e}^{-\mathrm{i} \pi \protect \hat {S}_x}$ at time $t/2$ which inverts the spin, sending state $\ket{m}$ to $\ket{-m}$. This transformation can be realized with a sequence of $S+1/2$ (for half-integer $S$) resonant $\pi$-rotations along $x$ or $y$, coupling pairs of $|m\rangle \leftrightarrow |-m\rangle$ states, followed by $Y_m(\pm \pi)$ pulses to rearrange populations. For instance, in the case of a spin $5/2$, $U_{\mathrm{echo}}$ is obtained by three independent rotations $|5/2\rangle \leftrightarrow |-5/2\rangle$, $|3/2\rangle \leftrightarrow |-3/2\rangle$ and $|1/2\rangle \leftrightarrow |-1/2\rangle$. Due to the lack of a direct matrix element between $\ket{m}$ and $\ket{-m}$ in the architecture considered here, each of these $\Delta m > 1$ transformations needs to be decomposed into $\Delta m = \pm 1$ transitions. This is done, for instance, using the strategy discussed in `Methods-Pulse sequences'. \\

  {\it Detection.} 
To realize the error detection, two additional ingredients are introduced in the system considered until now. The first one is a weak coupling of the qudit to a spin $s_A=1/2$ ancilla. The ancilla is described by adding to the Hamiltonian of Eq. \eqref{eq:Htot} the terms
\begin{equation}\label{eq:ancilla}
    \Omega_A \hat{s}_A^z + \sum_{k=x,y,z} \mathbb{J}_k \hat{s}_A^k \hat{S}^k,
\end{equation}
where $\{\hat{s}_A^x, \hat{s}_A^y, \hat{s}_A^z\}$ are spin-1/2 operators for the ancilla.
The first term in Eq. \eqref{eq:ancilla} describes the Zeeman coupling of the ancilla to the static magnetic field whilst the second one describes the ancilla-qubit coupling parametrised by the tensor $\mathbb{J}$. For $\mathbb{J}_{x,y} \ll |\Omega-\Omega_A|$, such that states of qudit and ancilla remain essentially factorised, the excitation energies of the ancilla $\Delta_A^{(m)}$ depend on the state $\ket{m}$ of the qudit via the diagonal coupling $\mathbb{J}_z \hat{s}_A^z \hat{S}_z$ only, i.e.
\begin{equation}
\Delta_A^{(m)} = g_A \mu_\mathrm{B} B_z + \mathbb{J}_z m.
\end{equation}
By irradiating the ancilla with a resonant magnetic pulse at angular frequency $\Delta_A^{(m)}$ it is thus possible to selectively excite the ancilla only if the qudit is in state $\ket{m}$. A subsequent measurement of the state of the ancilla then reveals whether the qudit state has support on $\ket{m}$ or not. This mechanism will be exploited in the following to detect the different possible errors without corrupting $\alpha$ and $\beta$. Apart from this selective excitation immediately followed by a measurement, the ancilla is always in its ground state, and thus it does not affect the previously developed treatment of the qudit incoherent dynamics. For this reason, its coupling to the nuclear spins is also irrelevant for the present discussion. 

  The second ingredient is a coupling of the magnetic molecule to a microwave resonator. Crucial steps towards achieving the strong coupling between magnetic molecules and a resonator have been experimentally demonstrated recently \cite{Gimeno2020}. This coupling can then be exploited to measure the ancilla, building on techniques well developed in the field of circuit quantum-electrodynamics \cite{Blais2004,Blais2021,Krantz2019} and adapted to Molecular Nanomagnets \cite{Jenkins2016,Carretta2021}. The coupling of the molecule to the resonator induces a shift of the resonance frequency of the resonator which depends on the ancilla-qudit state. As explained below, this can be exploited to measure the state of the ancilla without collapsing the qudit state.  

 The error detection is described in an abstract setting by a projective measurement with the projector operators given in Eq. \eqref{eq:projectors} of `Methods'. Given the difficulty to implement similar operators, that project into complex superpositions of system eigenstates, the detection is divided into two steps. In the first step, a sequence of pulses is performed which rotates the full basis of error words into the basis of $\ket{m}$ states in both error spaces corresponding to $\ket{0_L}$ and $\ket{1_L}$ (see panel `basis rotation' of Fig. \ref{fig:pulses} for $S=5/2$)  \cite{Chiesa2020}. This operation thus converts the detection of the projectors \eqref{eq:projectors} into an easier-to-implement measurement in the $\ket{m}$ basis, and the unitarity of the transformation ensures that $\alpha$ and $\beta$ are preserved. At this point, every possible post-error state is of the form $\alpha\ket{m}+\beta\ket{m'}$ for different pairs of states $(\ket{m},\ket{m'})$. 
 
We now aim to induce an excitation of the ancilla only for a superposition state of the qudit with components on $(\ket{m},\ket{m'})$, without collapsing the superposition. We achieve this by applying a two-tone two-photon drive at frequencies $\Delta_A^{(m)}$, $\Delta_A^{(m')}$ (panel `ancilla measurement' of Fig. \ref{fig:pulses}) \cite{Royer2018}. 
Then, in the dispersive limit, the coupling $G$ between resonator and ancilla induces a shift of the cavity angular frequency $\omega_c$ of $\pm G^2/\delta^m$, with $\delta^m=\Delta_A^{(m)}-\omega_c$ and the sign of the shift depending on the state of the ancilla \cite{Blais2004}. Since here we need to measure the ancilla irrespective of the state of the qudit in the subspace $(\ket{m},\ket{m'})$, a frequency-independent measurement of the state of the ancilla must be performed. Two different approaches to solve this same issue, by detecting the amplitude (but not the frequency) of the output field, are described in \cite{Royer2018}.
 The ancilla is then measured by exploiting its coupling to the resonator and the qudit wavefunction is projected onto such states.
The sequence of measurements is then repeated probing each (mutually exclusive) pair of $(\ket{m},\ket{m'})$ states sequentially, returning a yes/no answer at each step if the system is found in the corresponding error state, and stopping if a positive outcome is obtained. Hence, there will be at most $\lfloor S \rfloor$ measurements given that the number of possible errors for a qudit of spin $S$ is $\lceil S \rceil$, where $\lceil S \rceil$ ($\lfloor S \rfloor$) indicates the smallest integer larger (largest integer smaller) or equal than $S$. \\

  {\it Recovery.} After detection, the system has been projected into a superposition state of the form $\alpha \ket{m} + \beta \ket{m'}$ with known $m$ and $m'$. The simplest way to restore the encoded state $\alpha \ket{0_L} + \beta \ket{1_L}$ is then to first use a few $\pi$-pulses to send $\ket{m}\to \ket{1/2}$ and $\ket{m'}\to \ket{-1/2}$, and then to repeat the pulse sequence which implements the encoding (panel `recovery' in Fig. \ref{fig:pulses}). Alternatively, one can save a few pulses by redesigning the encoding sequence starting from each possible pair $\ket{m}$, $\ket{m'}$ resulting from detection. \\

  {\it Impact on performance.} The non-instantaneous duration of the QEC procedure in a realiztic implementation (during which information is {\it not protected}), together with related potential imperfections, may cause a loss of efficiency in the correction. We thus here discuss to what extent such effects can reduce the expected performance. 
  
   The operations described to implement the quantum error correction involve sequences of resonant pulses (and ancilla measurements) only. In electronic spin systems, a single $\pi$-pulse requires less than 10 ns for achieving a state transfer with high fidelity, and this time could be further reduced by pulse-shaping techniques \cite{Motzoi2009,Theis2018,Werninghaus2021}. The measurement time for the ancilla readout through a microwave resonator can be roughly estimated from the field of circuit QED to be of 40-100 ns with fidelity above 0.98 \cite{Walter2017,Blais2021}. Then, for the spins $S\le 9/2$ considered here the QEC procedure requires a total time ranging from a few hundreds of nanoseconds to few microseconds at most, and is hence much shorter than the decay time that can be allowed by the optimised code-words while ensuring a recovery fidelity above 0.99. Indeed, the latter can be of tens of microseconds, as visible from Fig. \ref{fig:opt1} and \ref{fig:opt2}.
   
   The practical implementation of the QEC is thus expected not to significantly reduce the correction performance for the qudits studied here. However, one could also predict that the growth of the complexity of the implementation for very large spins will eventually set a tradeoff between gain and duration of the QEC favouring the use of moderately large spins, similarly to what observed for spin-binomial codes \cite{Chiesa2020}.
Importantly, this limitation can be mitigated in the present scheme by optimising the code words at larger times. Moreover, it should be noted that the bottleneck in the specific experimental implementation proposed here for large spins is related to the rapid scaling of the number of pulses required with $S$. This, in turn, is due to the low connectivity of the $2S+1$ spin levels that permits resonant state transfers only between states with $\Delta m = \pm 1$. A possible way around this problem is to consider magnetic molecules with competing interactions \cite{vanSlageren2006,Schnack2010, Adelnia2015, Ghirri2015, Baker2016}, for which the multi-level structure used for the qudit encoding is given by low-energy states belonging to different multiplets that can provide larger state connectivity. \\
  
\noindent {\bf DISCUSSION}\vspace{0.1cm}

\noindent We have investigated decoherence effects produced by a realiztic nuclear spin bath on a spin qudit $S>1/2$ in Molecular Nanomagnets, by simulating the coupled system-bath dynamics via a cluster-correlation expansion. Building on this analysis, we have developed a versatile numerical strategy to construct optimal quantum error correction codes tailored for the specific spin-dephasing errors induced by the bath, thus bridging the gap between traditional general-purpose correction algorithms and the necessity of hardware-specific strategies meeting current experimental capabilities. The resulting qudit codes achieve a remarkable performance, and can be optimised by taking into account constraints on the time interval between subsequent QECs. Moreover, the increase in performance with the increase of the qudit spin is striking, signalling that the codes exploit the available levels of the molecular system as a resource very efficiently. The proposed codes can be implemented experimentally using standard sequences of resonant control pulses. Such sequences are explicitly designed and discussed, and their practical realization is predicted not to set important limits on the efficiency of the codes. Given these results, the proposed codes are a promising candidate for realizing error-protected quantum computational units embedded at the single molecule level, a central building block for implementing reliable quantum information processing on short-term molecular devices.
 
 Recent works \cite{Chen2020} point out that the CCE method used here correctly reproduces the phenomenology of coherence enhancement due to the existence of a nuclear diffusion barrier \cite{Graham2017}.  
An interesting perspective is the integration of the framework developed in this work with chemical engineering techniques for achieving an even longer lifetime of the error-corrected logical qubit through this mechanism. The synergy of tailored quantum error correction codes as investigated here with engineered nuclear spin distributions may pave the way towards a class of highly coherent molecular qubits.

 The framework developed in this work is general, and can be applied to a wide landscape of molecular systems and also to other individual spin systems such as impurities in semiconductors, in order to design a proof-of-principle experiment to demonstrate the effectiveness of the QEC code. In addition, it can be extended, in the future, to investigate decoherence effects affecting superpositions of spin states belonging to different spin multiplets \cite{vanSlageren2006,Schnack2010}. This would be interesting since, on the one hand, the use of many low-$m$ spin states belonging to different multiplets may allow one to increase the number of levels available for error correction without exasperating dephasing effects given by large $m-m'$ transitions. On the other hand, it would enable a thorough study of standard block encodings embedded in a single molecule, wherein a register of qubits is achieved through many effective spin-1/2 systems selected from different spin multiplets.  \\


\noindent {\bf METHODS}\vspace{0.1cm} \\
\noindent \textbf{Derivation of the effective Hamiltonian}.
The spin Hamiltonian describing the interacting evolution of the molecular spin $\mathcal{S}$ and the bath $\mathcal{B}$ of $N$ nuclear spins is
\begin{equation}\label{eq:Htot}
\hat{H}_{\mathcal{S}\mathcal{B}} = \hat{H}_{\mathcal{S}} + \sum_{n=1}^N \omega_n \hat{I}_n^z + \sum_{n=1}^N \hat{\bm{S}} \cdot \mathbb{D}_{n} \cdot \hat{\bm{I}}_n + \sum_{n\ne m} \hat{\bm{I}}_n \cdot \mathbb{E}_{n,m}\cdot \hat{\bm{I}}_m,
\end{equation}
where $\hat{H}_{\mathcal{S}}= D \hat{S}_z^2 + \Omega \hat{S}_z$, $\bm{\hat{S}} = \{\hat{S}^z, \hat{S}^+, \hat{S}^- \}$, $\bm{\hat{I}}_n = \{\hat{I}_n^z, \hat{I}_n^+, \hat{I}_n^- \}$. The tensors $\mathbb{D}_n$ contain dipole-dipole couplings between $\mathcal{S}$ and $\mathcal{B}$, while the tensors $\mathbb{E}_{n,m}$ contain nuclear-nuclear dipolar couplings. The elements of $\mathbb{D}_n$ satisfy 
\begin{equation}
\mathbb{D}_n^{\scriptscriptstyle ++} = (\mathbb{D}_n^{\scriptscriptstyle --})^*, \quad \mathbb{D}_n^{\scriptscriptstyle +-} = (\mathbb{D}_n^{\scriptscriptstyle +-})^* = \mathbb{D}_n^{\scriptscriptstyle -+}, \mathbb{D}_n^{+z} = (\mathbb{D}_n^{-z})^*, 
\end{equation} 
and the same holds for the corresponding elements of $\mathbb{E}_{n,m}$.
A canonical perturbative (Schrieffer-Wolff) transformation generated by 
\begin{equation}
G=\sum_{\beta=\pm,z} [G_{k}^{+\beta} \hat{S}^+ \hat{I}_n^\beta - \text{h.c.}],
\end{equation}
with h.c. indicating the hermitian conjugate, is constructed such that $\hat{H}_G = \mathrm{e}^{G} \hat{H} \mathrm{e}^{-G}$, within first order in $\Omega^{-1}$, does not contain off-diagonal couplings between $\mathcal{S}$ and $\mathcal{B}$ with respect to the states $\ket{m}$ \cite{Yao2006,Coish2008,Troiani2008,Bravyi2011}. As detailed in the following, $G$ is proportional to $\Omega^{-1}$ and leading orders in $\hat{H}_G$ can thus be computed from the Baker-Campbell-Hausdorf expansion
 \begin{equation}\label{eq:bkh}
     \hat{H}_G = \hat{H} + [G,\hat{H}] + \frac{1}{2!}[G,[G,\hat{H}]] + \dots \ .
 \end{equation}
 The coefficients $G_k^{\alpha\beta}$ are determined explicitly by imposing the cancellation of the off-diagonal couplings between $\mathcal{S}$ and $\mathcal{B}$ to first order. This results in the relation
 \begin{equation}
     [G,\hat{H}_{\mathcal{S}} + \sum_{n=1}^N \omega_n \hat{I}_n^z] = - \sum_{\substack{\alpha=\pm,\\ \beta=\pm,z}}\sum_{n=1}^N \mathbb{D}_n^{ \alpha\beta} \hat{S}^\alpha \hat{I}_n^\beta.
 \end{equation}
 The coefficients $G_k^{\alpha\beta}$ then read
 \begin{align}
    & G_k^{\scriptscriptstyle ++} = \frac{\mathbb{D}_k^{\scriptscriptstyle ++}}{\Omega +D + \omega_k }, \quad  G_k^{\scriptscriptstyle +-} = \frac{\mathbb{D}_k^{\scriptscriptstyle +-}}{\Omega+D - \omega_k } , \nonumber \\
 & G_k^{\scriptscriptstyle +z} = \frac{\mathbb{D}_k^{\scriptscriptstyle +z}}{\Omega +D }.
 \end{align}
These expressions are indeed of order $\Omega^{-1}$ and the transformation generated by $G$ is thus perturbative, such that its effect on the initial factorized qudit-nuclei state is neglected.
By keeping terms in Eq. \eqref{eq:bkh} to first order in $\Omega^{-1}$ only and neglecting energy-non-conserving terms, the effective Hamiltonian of Eq.s \eqref{eq:ham_eff} and \eqref{eq:ham_det} is obtained with coefficients
\begin{align}
a_n^{(0)} = &  \omega_n, \qquad b_n^{(0)} = 0, \qquad c_{n,m}^{(0)} = \mathbb{E}_{n,m}^{\scriptscriptstyle +-}, \nonumber \\
d_{n,m}^{(0)} = & \mathbb{E}_{n,m}^{zz}/2, \qquad a_n^{(1)}  = \mathbb{D}_n^{\scriptsize zz}, \nonumber \\
b_n^{(1)}  =  & \frac{2}{\Omega}  \left[ |\mathbb{D}_n^{\scriptstyle +z}|^2 - |\mathbb{D}_n^{\scriptscriptstyle{++}}|^2 - (\mathbb{D}_n^{\scriptscriptstyle{+-}})^2 \right], \nonumber \\
c_{n,m}^{(1)} = & \frac{2}{\Omega} (\mathbb{D}_n^{\scriptscriptstyle ++} \mathbb{D}_m^{\scriptscriptstyle --} + \mathbb{D}_n^{\scriptscriptstyle -+} \mathbb{D}_m^{\scriptscriptstyle +-}), \nonumber \\
d_{n,m}^{(1)} = & \frac{2}{\Omega}  \mathbb{D}_n^{+z} \mathbb{D}_m^{-z},\nonumber \\
a_n^{(2)} =  & \frac{2}{\Omega} \left[ |\mathbb{D}_n^{\scriptscriptstyle ++}|^2 - (\mathbb{D}_n^{\scriptscriptstyle +-})^2 \right] ,
\end{align}
and $b_n^{(2)} = c_n^{(2)} = d_n^{(2)} = 0$. The energy of $\mathcal{S}$ is also renormalized according to 
\begin{equation}
\tilde{\Omega} = \Omega + \frac{2 I(I+1)}{\Omega} \sum_{n=1}^N \left[|\mathbb{D}_n^{\scriptscriptstyle ++}|^2 + (\mathbb{D}_n^{\scriptscriptstyle +-})^2 \right], 
\end{equation}
but this is absorbed into the interaction picture in Eq. \eqref{eq:ham_eff}.\\

\noindent \textbf{Simulations and dephasing timescales.} The configuration of nuclear spins in space is generated randomly within a sphere of radius $15 \, \AA$ from the spin $\mathcal{S}$, as sketched in Fig. \ref{fig0}. Further, a minimal distance of $3 \,\AA$ is considered between nuclear spins and between each nuclear spin and $\mathcal{S}$. In all the simulations presented in this work, configurations of $N=100$ nuclear spins are considered, whose initial state is taken to be thermal at temperature $T=2$ K. Moreover, a static magnetic field $B_z=1$ T along $z$ is assumed, for achieving the regime of large Zeeman energy of $\mathcal{S}$.
The decoherence function,
\begin{equation}
L_{nm}(t)=\text{tr}_\mathcal{B} \left[ \mathrm{e}^{- \mathrm{i} \hat{H}_{\mathcal{B},n} t} \rho_\mathcal{B}(0) \mathrm{e}^{\mathrm{i} \hat{H}_{\mathcal{B},m} t} \right] \ ,
 \end{equation} 
 is computed by means of a cluster-correlation expansion (CCE) \cite{Yang2008,Yang2009}.
This expansion decomposes $L_{nm}(t)$ as a product of contributions originating from clusters involving an increasing number of nuclear spins, and is formally equivalent to a perturbative expansion in small intra-bath effective couplings. Clusters involving more and more spins contribute smaller and smaller corrections to $L_{nm}(t)$, justifying a truncation of the expansion to few-spin clusters for practical applications. For inclusion up to $n$-size clusters, we call this truncation CCE-$n$, which yields the truncated function $L^{(n)}_{nm}(t)$.

 The effect of inhomogeneous broadening is well captured by CCE-1 \cite{Yang2009,Maze2008}. Given that nuclear gaps are of the order of millikelvin in magnitude, an initial thermal state of the nuclear spin-bath at temperatures $T$ of a few kelvins is to a good approximation a fully unpolarized state 
\begin{equation}
\rho_{\mathcal{B}}(0) = \frac{\mathrm{e}^{-H_\mathcal{B}^{(0)}/k_B T}}{\text{tr}_\mathcal{B} [ \mathrm{e}^{-H_\mathcal{B}^{(0)}/k_B T}] }\simeq \mathbb{1}_{\mathcal{B}}/2^N.\end{equation}
 Under this approximation the CCE-1 can be solved analytically also for $S>1/2$ giving 
\begin{equation}
L_{nm}^{(1)}(t) = \prod_{k=1}^N \cos\left[\frac{(n-m) \mathbb{D}_k^{zz} }{2} t \right]. 
\end{equation}
Here, $\mathbb{D}_k^{zz}$ is the hyperfine coupling $\propto\hat{S}_z \hat{I}_k^z$ between $\mathcal{S}$ and the $k$-th nuclear spin.
For small $\mathbb{D}_k^{zz} t$, this is well approximated by $\mathrm{e}^{-(n-m)^2 \Gamma^2 t^2}$ with $\Gamma^2 = \sum_k (D_k^{zz})^2/4$ as given in the main text.  

 For the echo dynamics, we find that CCE-2 gives essentially converged results, as tested by including also larger clusters. For this reason, the numerical results reported here are obtained using CCE-2. This convergence confirms that intrinsic nuclear flip-flop processes give the dominant contribution to spin dephasing after echo. Indeed, inspection of the magnitude of the coefficients $a_{n,m}^{(k)}$, $b_{n,m}^{(k)}$, $c_{n,m}^{(k)}$, $d_{n,m}^{(k)}$ reveals three fundamental interaction scales at play, which, ordered by decreasing strength, are associated to \emph{(i)} the diagonal coupling between $\mathcal{S}$ and nuclear spins (terms $\propto \hat{S}_z \hat{I}_k^z$ which are compensated for by the echo),  
 \emph{(ii)} the intrinsic interacting evolution (terms $\propto \hat{I}_n^{\alpha} \hat{I}_m^{\beta}$), \emph{(iii)} the  $\mathcal{S}$-conditioned interacting evolution (terms $\propto \hat{S}_z \hat{I}_m^\alpha \hat{I}_n^\beta$). These different energy scales are responsible for contributions to decoherence over different timescales, with {\it (ii)} being dominant in the echo decay.  \\
 
\noindent \textbf{Nuclear spin qudit.}  \label{app:nuclear}
The quantitative analyses presented in this work are focused on the case of an electronic spin qudit. Nevertheless, the theoretical framework applies also to the case of a nuclear spin qudit. The crucial approximation underlying the physical model studied here is that the qudit energy gaps are much larger than the gaps of surrounding nuclear spins of the bath. For an electronic qudit, these energy differences can be made intrinsically large by using a sufficiently large static magnetic field. For a nuclear spin qudit of a magnetic ion (coupled to an electronic spin), whose interaction with surrounding nuclear spins is mediated by virtual excitations of the electronic spin, the necessary energy difference mainly originates from contact hyperfine interaction between the nuclear and electronic spin. The construction of the effective 
Hamiltonian, and the hierarchy of the interaction scales at play, then follows as discussed above. \\

\noindent \textbf{Derivation of numerical qudit codes.} \label{app:numcodes}
The iteration defined by Eq. \eqref{eq:iteration} is first used to determine the error operators given in Eq. \eqref{eq:err_ansatz}. 
The numerical code words $\ket{0_L}$ and $\ket{1_L}$ are found by starting from the following \emph{ansatz}, inspired by spin-binomial codes \cite{Chiesa2020},
\begin{equation}
\ket{0_L/1_L} = \sum_{\stackrel{\ell=0}{\ell \textrm{ odd/even}}}^{S+1/2} \gamma_\ell^{(0)/(1)} \ket{-S + \ell}.
\end{equation}
This \emph{ansatz} permits one to impose Eq. \eqref{eq:kl2} by construction and hence to reduce the number of free parameters, thanks to $\ket{0_L}$ and $\ket{1_L}$ having non-zero components on different sets of $\ket{m}$ states. 

Knill-Laflamme conditions \eqref{eq:kl} are finally enforced on the coefficients $\gamma_{\ell}^{(0)/(1)}$ by numerically minimizing the function
\begin{equation}
\sum_{k,j=0}^{2S} \left\lvert \bra{0_L} \hat{E}_k^\dagger \hat{E}_j \ket{0_L} -\bra{1_L} \hat{E}_k^\dagger \hat{E}_j \ket{1_L}  \right\lvert.
\end{equation}

\noindent \textbf{Detection and recovery.} \label{app:qec}
The abstract detection and recovery operations follow from the general quantum error correction theory of Ref. \cite{Knill1997}. Once a set of operators $\{\hat{E}_k\}_{k=0,\dots, \lfloor S \rfloor}$ to be corrected with a spin $S$ is identified, the two error subspaces corresponding to $\ket{0_L}$ and $\ket{1_L}$ are first determined. These two subspace are defined as the linear span of the states $\hat{E}_k\ket{0_L}$ and $\hat{E}_k\ket{1_L}$ for all $k$, respectively. For each of these subspaces, a basis $\{\ket{e_k^{(0)}}\}$ ($\{\ket{e_k^{(1)}}\}$) is selected. This is chosen here to be given by a Gram-Schmidt orthonormalization of states $\hat{E}_k\ket{0_L}$ ($\hat{E}_k\ket{1_L}$). The detection measurement is then described by the projectors 
\begin{equation}\label{eq:projectors}
\hat{P}_k = \ket{e_k^{(0)}}\bra{e_k^{(0)} }+ \ket{e_k^{(1)}}\bra{e_k^{(1)}},
\end{equation}
with $k=0,\ldots,\lfloor S \rfloor$. Finally, the recovery operation, given an outcome $j$ of the measurement, maps back the states $\ket{e_j^{(0)}}$ and $\ket{e_j^{(1)}}$ corresponding to that outcome to the computational states $\ket{0_L}$ and $\ket{1_L}$, respectively. Since the coefficients $\alpha$ and $\beta$ of the encoded state have been preserved, this operation then fully restores the logical state $\ket{\psi_L}$. The recovery is formalised by a set of transformations $\{\hat{R}_k\}$ such that 
 \begin{equation}
 \hat{R}_k \ket{e_k^{(c)}} = \ket{c_L},
 \end{equation}
 for $c=0,1$. Here, each transformation $\hat{R}_k$ is constructed as a rotation in the two-dimensional space spanned by $\ket{e_k^{(c)}}$ and $\ket{c_L}$. \\

\noindent {\bf Pulse sequences}. To systematically convert a generic transformation $U$ acting on the state space of a spin $S>1/2$ into a sequence of resonant $Y_m(\theta)$ pulses, one can exploit known decomposition strategies from quantum control theory \cite{Dalessandro2007, Chiesa2020}. In a first step, the unitary $U$ can be decomposed into a sequence of two-state planar rotations using the algorithm given in Ref. \cite{Dalessandro2007}. This gives a sequence of two-state transfers which involve states $\ket{m}$ and $\ket{m'}$ also with $|m-m'|>1$. To further convert such two-state rotations into rotations $Y_m(\theta)$, {\it i.e.} with $|m-m'|=1$, one finally exploits $\pi$-pulses to bring the population of $m'$ close to $m$ and back. For instance, defining
\begin{equation}
    Y_{m,m'}(\theta) = \exp\big[\theta/2 (-\ket{m}\bra{m'} +\ket{m'}\bra{m} \big]
\end{equation}
for $m'>m$, one can iteratively exploit the properties 
\begin{align}
 Y_{m,m+2} (\theta) & =  Y_{m,m+1}(\pi) Y_{m+1,m+2} (-\theta) Y_{m,m+1}(-\pi). \\
& =  Y_{m,m+1}(-\pi) Y_{m+1,m+2} (\theta) Y_{m,m+1}(\pi).
\end{align}
As an example, the pulse sequence depicted in panel `basis rotation' of Fig. \ref{fig:pulses}a, which realizes a transformation mapping the basis of error words $\ket{e_k^{(c)}}$ to the basis of $\ket{m}$ states, is obtained with the procedure sketched here. The resulting angles are given in Table \ref{tab:1}.\\
 \begin{table}
 \caption{ Angles for the pulse sequence realizing the QEC for $S=5/2$ as depicted in Fig. \ref{fig:pulses}.
}
 \label{tab:1}
 \begin{ruledtabular}
\begin{tabular}{p{0.20\linewidth} p{0.20\linewidth} | p{0.20\linewidth} p{0.20\linewidth}}
\multicolumn{4}{c}{\bf Angles for the pulses in Fig. 6}\\
\hline 
\multicolumn{2}{c |}{$\theta$ angles} & \multicolumn{2}{c}{$\phi$ angles} \\
\hline
\hline
$\theta_1$ & 0.342 $\pi$ & $\phi_2$ & 0.339 $\pi$ \\
$\theta_2$ & 0.560 $\pi$  & $\phi_2$ & 0.562 $\pi$ \\
$\theta_3$ & 0.552 $\pi$ & $\phi_3$ & 0.888 $\pi$ \\
$\theta_4$ & $-$0.085 $\pi$ & $\phi_4$ & $-$0.440 $\pi$ \\
$\theta_5$ & 1.531 $\pi$ & $\phi_5$ & 1.538 $\pi$ \\
\end{tabular}
\end{ruledtabular}
\end{table}

\begin{figure}
	\centering
	\includegraphics[width=\linewidth]{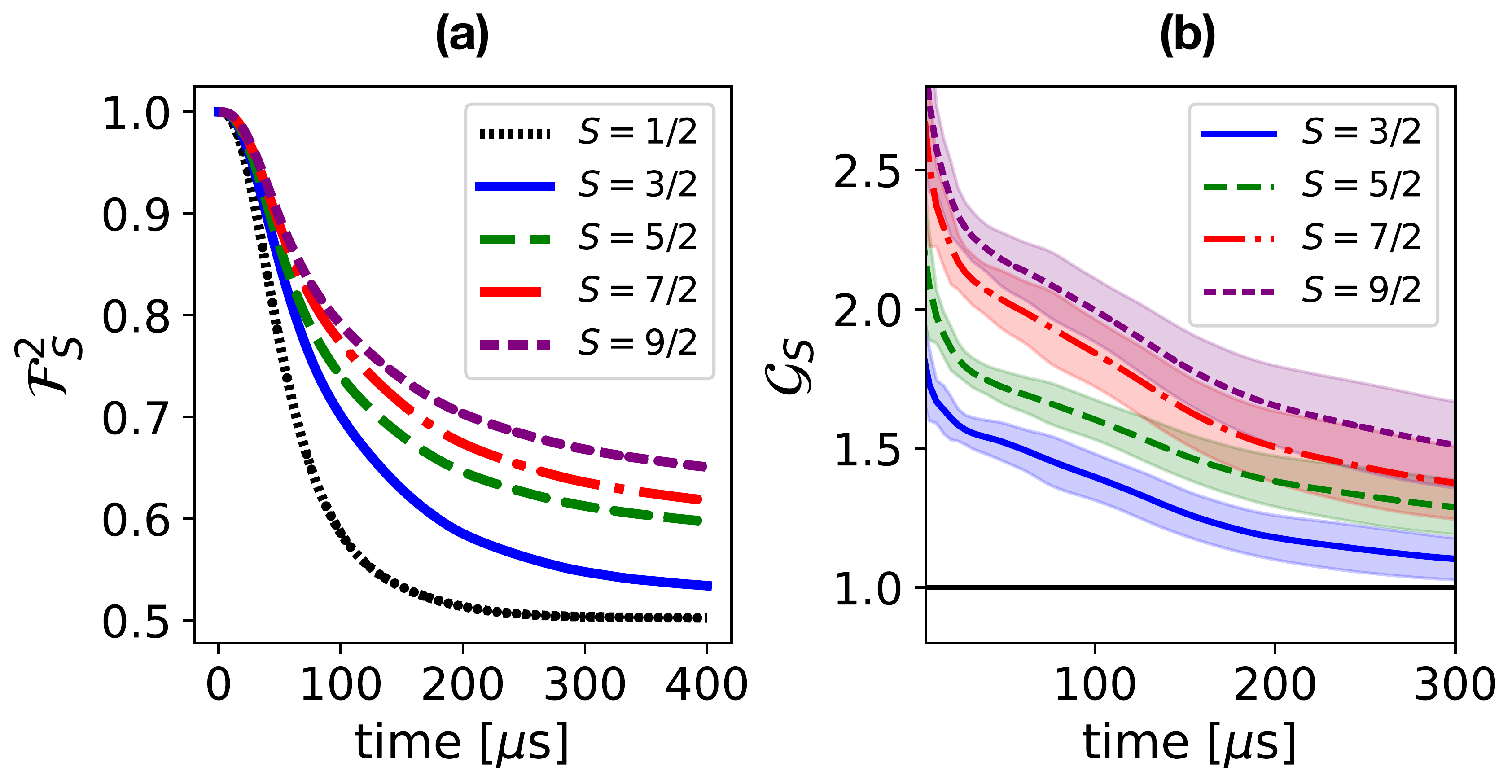}
	\caption{{\bf Performance of spin-binomial codes}.  \textbf{(a)} Average squared fidelity as a function of time for different values of the qudit spin $S$, for initial state $\ket{\psi_L} = (\ket{0_L} + \mathrm{i} \ket{1_L})/\sqrt{2}$ and spin-binomial code words $\ket{0_L}$ and $\ket{1_L}$. The results are averaged over $2^6$ nuclear spin configurations. \textbf{(b)} Gain, defined in Eq. \eqref{eq:gain}, with respect to a spin 1/2, for the same data of panel (a). The shaded area are constructed as in Fig. \ref{fig:gainopt}.}
	\label{fig:gain_spin_binomial}
\end{figure}

\noindent \textbf{Spin-binomial codes.} \label{app:spinbin}
In order to compare the numerical codes with other qudit approaches to spin dephasing, we test recently-proposed spin-binomial codes \cite{Chiesa2020}. These codes are based on a description of spin dephasing as produced by a Markovian bath which couples to the system via operator $\sqrt{\gamma} \hat{S}_z$. The latter model can only describe an exponential decay of coherence with rate $\gamma$ and contributions of order $(\gamma t)^n$ to the density matrix can be computed exactly and are determined by powers up to $\hat{S}_z^n$. 
We find that, while spin-binomial codes still give interesting performance, they are largely overwhelmed by the numerical codes. This can be appreciated by comparing Fig. \ref{fig:opt1} and \ref{fig:gainopt} with Fig. \ref{fig:gain_spin_binomial}, both in terms of fidelity and gain. 
The fact that spin-binomial codes still give an increasing gain for increasing qudit spin despite being designed for a simpler dephasing mechanism, suggests that small powers of the coupling operators $\hat{S}_z$ play a fundamental role in the decoherence process also in the present scenario, over short timescales with respect to the intra-bath interaction strength. \\

 \noindent{\bf \small \small DATA AVAILABILITY}\vspace{0.1cm}
 
\noindent The datasets generated during and/or analysed during the current study are available from the corresponding author on reasonable request.\\

\noindent {\bf \small CODE AVAILABILITY}\vspace{0.1cm}
 
\noindent The codes developed for the numerical simulations are available from the corresponding author on reasonable request. \\

\noindent {\bf \small ACKNOWLEDGMENTS} \vspace{0.1cm}
 
\noindent This work received financial support from the European Project `Scaling Up quantum computation with MOlecular spins' (SUMO) of the call QuantERA, cofunded by the Italian MUR, and the European Union's Horizon 2020 program under Grant Agreement No. 862893 (FET-OPEN project FATMOLS).\\

\noindent {\bf \small AUTHOR CONTRIBUTIONS}\vspace{0.1cm}
 
\noindent S.W., P.S. and S.C. conceived the research. F.P. derived the theoretical model and produced the numerical data. F.P., A.C. and S.C. analyzed the data. All authors discussed the results. F.P. and A.C. wrote the manuscript, with input from all the co-authors.\\

\noindent {\bf \small COMPETING INTERESTS}\vspace{0.1cm}
 
\noindent The authors declare no competing interests. \\


\vspace{-2cm}
\begin{thebibliography}{10}
\expandafter\ifx\csname url\endcsname\relax
  \def\url#1{\texttt{#1}}\fi
\expandafter\ifx\csname urlprefix\endcsname\relax\def\urlprefix{URL }\fi
\providecommand{\bibinfo}[2]{#2}
\providecommand{\eprint}[2][]{\url{#2}}

\bibitem{Terhal2015}
\bibinfo{author}{Terhal, B.~M.}
\newblock \bibinfo{title}{Quantum error correction for quantum memories}.
\newblock \emph{\bibinfo{journal}{Rev. Mod. Phys.}}
  \textbf{\bibinfo{volume}{87}}, \bibinfo{pages}{307--346}
  (\bibinfo{year}{2015}).
\newblock \urlprefix\url{https://link.aps.org/doi/10.1103/RevModPhys.87.307}.

\bibitem{Preskill2018}
\bibinfo{author}{Preskill, J.}
\newblock \bibinfo{title}{{Quantum Computing in the NISQ era and beyond}}.
\newblock \emph{\bibinfo{journal}{Quantum}} \textbf{\bibinfo{volume}{2}},
  \bibinfo{pages}{79} (\bibinfo{year}{2018}).
\newblock \urlprefix\url{https://quantum-journal.org/papers/q-2018-08-06-79/}.

\bibitem{Tacchino2020}
\bibinfo{author}{Tacchino, F.}, \bibinfo{author}{Chiesa, A.},
  \bibinfo{author}{Carretta, S.} \& \bibinfo{author}{Gerace, D.}
\newblock \bibinfo{title}{Quantum computers as universal quantum simulators:
  state-of-the-art and perspectives}.
\newblock \emph{\bibinfo{journal}{Adv. Quantum Technol.}}
  \textbf{\bibinfo{volume}{3}}, \bibinfo{pages}{1900052}
  (\bibinfo{year}{2020}).
\newblock
  \urlprefix\url{https://onlinelibrary.wiley.com/doi/abs/10.1002/qute.201900052}.

\bibitem{NielsenChuang2000}
\bibinfo{author}{Nielsen, M.~A.} \& \bibinfo{author}{Chuang, I.~L.}
\newblock \emph{\bibinfo{title}{{Quantum Computation and Quantum Information}}}
  (\bibinfo{publisher}{Cambridge University Press}, \bibinfo{year}{2000}).

\bibitem{Gottesman2001}
\bibinfo{author}{Gottesman, D.}, \bibinfo{author}{Kitaev, A.} \&
  \bibinfo{author}{Preskill, J.}
\newblock \bibinfo{title}{Encoding a qubit in an oscillator}.
\newblock \emph{\bibinfo{journal}{Phys. Rev. A}} \textbf{\bibinfo{volume}{64}},
  \bibinfo{pages}{012310} (\bibinfo{year}{2001}).
\newblock \urlprefix\url{https://link.aps.org/doi/10.1103/PhysRevA.64.012310}.

\bibitem{Troiani2012}
\bibinfo{author}{Troiani, F.}, \bibinfo{author}{Bellini, V.} \&
  \bibinfo{author}{Affronte, M.}
\newblock \bibinfo{title}{Decoherence induced by hyperfine interactions with
  nuclear spins in antiferromagnetic molecular rings}.
\newblock \emph{\bibinfo{journal}{Phys. Rev. B}} \textbf{\bibinfo{volume}{77}},
  \bibinfo{pages}{054428} (\bibinfo{year}{2008}).
\newblock \urlprefix\url{https://link.aps.org/doi/10.1103/PhysRevB.77.054428}.

\bibitem{Leghtas2013}
\bibinfo{author}{Leghtas, Z.}, \bibinfo{author}{Kirchmair, G.},
  \bibinfo{author}{Vlastakis, B.}, \bibinfo{author}{Schoelkopf, R.~J.},
  \bibinfo{author}{Devoret, M.~H.} \& \bibinfo{author}{Mirrahimi, M.}
\newblock \bibinfo{title}{Hardware-efficient autonomous quantum memory
  protection}.
\newblock \emph{\bibinfo{journal}{Phys. Rev. Lett.}}
  \textbf{\bibinfo{volume}{111}}, \bibinfo{pages}{120501}
  (\bibinfo{year}{2013}).
\newblock
  \urlprefix\url{https://link.aps.org/doi/10.1103/PhysRevLett.111.120501}.

\bibitem{Mirrahimi2014}
\bibinfo{author}{Mirrahimi, M.} \emph{et~al.}
\newblock \bibinfo{title}{Dynamically protected cat-qubits: a new paradigm for
  universal quantum computation}.
\newblock \emph{\bibinfo{journal}{New J. Phys.}} \textbf{\bibinfo{volume}{16}},
  \bibinfo{pages}{045014} (\bibinfo{year}{2014}).
\newblock
  \urlprefix\url{https://doi.org/10.1088%2F1367-2630%2F16%2F4%2F045014}.

\bibitem{Vlastakis2013}
\bibinfo{author}{Vlastakis, B.} \emph{et~al.}
\newblock \bibinfo{title}{Deterministically encoding quantum information using
  100-photon {S}chr{\"o}dinger cat states}.
\newblock \emph{\bibinfo{journal}{Science}} \textbf{\bibinfo{volume}{342}},
  \bibinfo{pages}{607--610} (\bibinfo{year}{2013}).
\newblock \urlprefix\url{https://science.sciencemag.org/content/342/6158/607}.

\bibitem{Linshu2017}
\bibinfo{author}{Li, L.}, \bibinfo{author}{Zou, C.-L.},
  \bibinfo{author}{Albert, V.~V.}, \bibinfo{author}{Muralidharan, S.},
  \bibinfo{author}{Girvin, S.~M.} \& \bibinfo{author}{Jiang, L.}
\newblock \bibinfo{title}{Cat codes with optimal decoherence suppression for a
  lossy bosonic channel}.
\newblock \emph{\bibinfo{journal}{Phys. Rev. Lett.}}
  \textbf{\bibinfo{volume}{119}}, \bibinfo{pages}{030502}
  (\bibinfo{year}{2017}).
\newblock
  \urlprefix\url{https://link.aps.org/doi/10.1103/PhysRevLett.119.030502}.

\bibitem{Hu2019}
\bibinfo{author}{Hu, L.} \emph{et~al.}
\newblock \bibinfo{title}{Quantum error correction and universal gate set
  operation on a binomial bosonic logical qubit}.
\newblock \emph{\bibinfo{journal}{Nat. Phys.}} \textbf{\bibinfo{volume}{15}},
  \bibinfo{pages}{503--508} (\bibinfo{year}{2019}).
\newblock
  \urlprefix\url{https://www.nature.com/articles/s41567-018-0414-3#citeas}.

\bibitem{Michael2016}
\bibinfo{author}{Michael, M.~H.} \emph{et~al.}
\newblock \bibinfo{title}{New class of quantum error-correcting codes for a
  bosonic mode}.
\newblock \emph{\bibinfo{journal}{Phys. Rev. X}} \textbf{\bibinfo{volume}{6}},
  \bibinfo{pages}{031006} (\bibinfo{year}{2016}).
\newblock \urlprefix\url{https://link.aps.org/doi/10.1103/PhysRevX.6.031006}.

\bibitem{Chiesa2020}
\bibinfo{author}{Chiesa, A.}, \bibinfo{author}{Macaluso, E.},
  \bibinfo{author}{Petiziol, F.}, \bibinfo{author}{Wimberger, S.},
  \bibinfo{author}{Santini, P.} \& \bibinfo{author}{Carretta, S.}
\newblock \bibinfo{title}{Molecular nanomagnets as qubits with embedded
  quantum-error correction}.
\newblock \emph{\bibinfo{journal}{J. Phys. Chem. Lett.}}
  \textbf{\bibinfo{volume}{11}}, \bibinfo{pages}{8610--8615}
  (\bibinfo{year}{2020}).
\newblock \urlprefix\url{https://doi.org/10.1021/acs.jpclett.0c02213}.

\bibitem{Pirandola2008}
\bibinfo{author}{Pirandola, S.}, \bibinfo{author}{Mancini, S.},
  \bibinfo{author}{Braunstein, S.~L.} \& \bibinfo{author}{Vitali, D.}
\newblock \bibinfo{title}{Minimal qudit code for a qubit in the phase-damping
  channel}.
\newblock \emph{\bibinfo{journal}{Phys. Rev. A}} \textbf{\bibinfo{volume}{77}},
  \bibinfo{pages}{032309} (\bibinfo{year}{2008}).
\newblock \urlprefix\url{https://link.aps.org/doi/10.1103/PhysRevA.77.032309}.

\bibitem{Cafaro2012}
\bibinfo{author}{Cafaro, C.}, \bibinfo{author}{Maiolini, F.} \&
  \bibinfo{author}{Mancini, S.}
\newblock \bibinfo{title}{Quantum stabilizer codes embedding qubits into
  qudits}.
\newblock \emph{\bibinfo{journal}{Phys. Rev. A}} \textbf{\bibinfo{volume}{86}},
  \bibinfo{pages}{022308} (\bibinfo{year}{2012}).
\newblock \urlprefix\url{https://link.aps.org/doi/10.1103/PhysRevA.86.022308}.

\bibitem{Hussain2018}
\bibinfo{author}{Hussain, R.} \emph{et~al.}
\newblock \bibinfo{title}{Coherent manipulation of a molecular {Ln}-based
  nuclear qudit coupled to an electron qubit}.
\newblock \emph{\bibinfo{journal}{J. Am. Chem. Soc.}}
  \textbf{\bibinfo{volume}{140}}, \bibinfo{pages}{9814--9818}
  (\bibinfo{year}{2018}).
\newblock \urlprefix\url{https://doi.org/10.1021/jacs.8b05934}.

\bibitem{Chiesa2021}
\bibinfo{author}{Chiesa, A.}, \bibinfo{author}{Petiziol, F.},
  \bibinfo{author}{Macaluso, E.}, \bibinfo{author}{Wimberger, S.},
  \bibinfo{author}{Santini, P.} \& \bibinfo{author}{Carretta, S.}
\newblock \bibinfo{title}{Embedded quantum-error correction and
  controlled-phase gate for molecular spin qubits}.
\newblock \emph{\bibinfo{journal}{AIP Adv.}} \textbf{\bibinfo{volume}{11}},
  \bibinfo{pages}{025134} (\bibinfo{year}{2021}).
\newblock \urlprefix\url{https://doi.org/10.1063/9.0000166}.

\bibitem{Grover}
\bibinfo{author}{Godfrin, C.} \emph{et~al.}
\newblock \bibinfo{title}{Operating quantum states in single magnetic
  molecules: implementation of {G}rover's quantum algorithm}.
\newblock \emph{\bibinfo{journal}{Phys. Rev. Lett.}}
  \textbf{\bibinfo{volume}{119}}, \bibinfo{pages}{187702}
  (\bibinfo{year}{2017}).
\newblock
  \urlprefix\url{https://link.aps.org/doi/10.1103/PhysRevLett.119.187702}.

\bibitem{Bader2014}
\bibinfo{author}{Bader, K.} \emph{et~al.}
\newblock \bibinfo{title}{Room temperature quantum coherence in a potential
  molecular quit}.
\newblock \emph{\bibinfo{journal}{Nat. Commun.}} \textbf{\bibinfo{volume}{5}},
  \bibinfo{pages}{5304--5309} (\bibinfo{year}{2014}).
\newblock \urlprefix\url{https://www.nature.com/articles/ncomms6304}.

\bibitem{Zadrozny}
\bibinfo{author}{Zadrozny, J.~M.}, \bibinfo{author}{Niklas, J.},
  \bibinfo{author}{Poluektov, O.~G.} \& \bibinfo{author}{Freedman, D.~E.}
\newblock \bibinfo{title}{Millisecond coherence time in a tunable molecular
  electronic spin qubit.}
\newblock \emph{\bibinfo{journal}{ACS Cent. Sci.}}
  \textbf{\bibinfo{volume}{1}}, \bibinfo{pages}{488} (\bibinfo{year}{2015}).
\newblock \urlprefix\url{J. M. Zadrozny, J. Niklas, O. G. Poluektov, and D. E.
  Freedman, ACS Cent. Sci. 1, 488 (2015)}.

\bibitem{Hill}
\bibinfo{author}{Shiddiq, M.}, \bibinfo{author}{Komijani, D.},
  \bibinfo{author}{Duan, Y.}, \bibinfo{author}{Gaita-Ari$\tilde{\text n}$o,
  A.}, \bibinfo{author}{Coronado, E.} \& \bibinfo{author}{Hill, S.}
\newblock \bibinfo{title}{Enhancing coherence in molecular spin qubits via
  atomic clock transitions.}
\newblock \emph{\bibinfo{journal}{Nature}} \textbf{\bibinfo{volume}{531}},
  \bibinfo{pages}{348--351} (\bibinfo{year}{2016}).
\newblock \urlprefix\url{https://pubmed.ncbi.nlm.nih.gov/26983539/}.

\bibitem{Atzori2016}
\bibinfo{author}{Atzori, M.}, \bibinfo{author}{Tesi, L.},
  \bibinfo{author}{Morra, E.}, \bibinfo{author}{Chiesa, M.},
  \bibinfo{author}{Sorace, L.} \& \bibinfo{author}{Sessoli, R.}
\newblock \bibinfo{title}{Room-temperature quantum coherence and rabi
  oscillations in vanadyl phthalocyanine: Toward multifunctional molecular spin
  qubits.}
\newblock \emph{\bibinfo{journal}{J. Am. Chem. Soc.}}
  \textbf{\bibinfo{volume}{138}}, \bibinfo{pages}{2154--2157}
  (\bibinfo{year}{2016}).
\newblock \urlprefix\url{https://pubs.acs.org/doi/10.1021/jacs.5b13408}.

\bibitem{Atzori2017}
\bibinfo{author}{Atzori, M.} \emph{et~al.}
\newblock \bibinfo{title}{Spin dynamics and low energy vibrations: Insights
  from vanadyl-based potential molecular qubits.}
\newblock \emph{\bibinfo{journal}{J. Am. Chem. Soc.}}
  \textbf{\bibinfo{volume}{139}}, \bibinfo{pages}{4338--4341}
  (\bibinfo{year}{2017}).
\newblock \urlprefix\url{https://pubs.acs.org/doi/10.1021/jacs.7b01266}.

\bibitem{Atzori2018}
\bibinfo{author}{Atzori, M.} \emph{et~al.}
\newblock \bibinfo{title}{Structural effects on the spin dynamics of potential
  molecular qubits.}
\newblock \emph{\bibinfo{journal}{Inorg. Chem.}} \textbf{\bibinfo{volume}{57}},
  \bibinfo{pages}{731--740} (\bibinfo{year}{2018}).
\newblock
  \urlprefix\url{https://pubs.acs.org/doi/abs/10.1021/acs.inorgchem.7b02616}.

\bibitem{Atzori_JACS}
\bibinfo{author}{Atzori, M.} \emph{et~al.}
\newblock \bibinfo{title}{Quantum coherence times enhancement in
  vanadium({IV})-based potential molecular qubits: the key role of the vanadyl
  moiety}.
\newblock \emph{\bibinfo{journal}{J. Am. Chem. Soc.}}
  \textbf{\bibinfo{volume}{138}}, \bibinfo{pages}{11234--11244}
  (\bibinfo{year}{2016}).
\newblock \urlprefix\url{https://pubs.acs.org/doi/10.1021/jacs.6b05574}.

\bibitem{Freedman_JACS}
\bibinfo{author}{Yu, C.-J.} \emph{et~al.}
\newblock \bibinfo{title}{Long coherence times in nuclear spin-free vanadyl
  qubits}.
\newblock \emph{\bibinfo{journal}{J. Am. Chem. Soc.}}
  \textbf{\bibinfo{volume}{138}}, \bibinfo{pages}{14678--14685}
  (\bibinfo{year}{2016}).
\newblock \urlprefix\url{https://pubs.acs.org/doi/abs/10.1021/jacs.6b08467}.

\bibitem{Freedman2014}
\bibinfo{author}{Graham, M.~J.} \emph{et~al.}
\newblock \bibinfo{title}{Influence of electronic spin and spin–orbit
  coupling on decoherence in mononuclear transition metal complexes.}
\newblock \emph{\bibinfo{journal}{J. Am. Chem. Soc.}}
  \textbf{\bibinfo{volume}{136}}, \bibinfo{pages}{7623--7626}
  (\bibinfo{year}{2014}).
\newblock \urlprefix\url{https://pubs.acs.org/doi/10.1021/ja5037397}.

\bibitem{Freedman_Ni}
\bibinfo{author}{Wojnar, M.~K.}, \bibinfo{author}{Laorenza, D.~W.},
  \bibinfo{author}{Schaller, R.~D.} \& \bibinfo{author}{Freedman, D.~E.}
\newblock \bibinfo{title}{Nickel({II}) metal complexes as optically addressable
  qubit candidates.}
\newblock \emph{\bibinfo{journal}{J. Am. Chem. Soc.}}
  \textbf{\bibinfo{volume}{142}}, \bibinfo{pages}{14826--14830}
  (\bibinfo{year}{2020}).
\newblock \urlprefix\url{https://pubs.acs.org/doi/10.1021/jacs.0c06909}.

\bibitem{Freedman_Cr}
\bibinfo{author}{Fataftah, M.}, \bibinfo{author}{Zadrozny, J.~M.},
  \bibinfo{author}{Coste, S.~C.}, \bibinfo{author}{Graham, M.~J.},
  \bibinfo{author}{Rogers, D.~M.} \& \bibinfo{author}{Freedman, D.~E.}
\newblock \bibinfo{title}{Employing forbidden transitions as qubits in a
  nuclear spin-free chromium complex.}
\newblock \emph{\bibinfo{journal}{J. Am. Chem. Soc.}}
  \textbf{\bibinfo{volume}{138}}, \bibinfo{pages}{1344} (\bibinfo{year}{2016}).
\newblock \urlprefix\url{https://pubs.acs.org/doi/10.1021/jacs.5b11802}.

\bibitem{Mn19powell}
\bibinfo{author}{Ako, A.~M.} \emph{et~al.}
\newblock \bibinfo{title}{A ferromagnetically coupled $\mathrm{Mn}_{19}$
  aggregate with a record {$S$}=83/2 ground spin state}.
\newblock \emph{\bibinfo{journal}{Angew. Chem. Int. Ed.}}
  \textbf{\bibinfo{volume}{45}}, \bibinfo{pages}{4926--4929}
  (\bibinfo{year}{2006}).
\newblock
  \urlprefix\url{https://onlinelibrary.wiley.com/doi/abs/10.1002/anie.200601467}.

\bibitem{giant}
\bibinfo{author}{Baniodeh, A.} \emph{et~al.}
\newblock \bibinfo{title}{High spin cycles: topping the spin record for a
  single molecule verging on quantum criticality.}
\newblock \emph{\bibinfo{journal}{npj Quantum Mater.}}
  \textbf{\bibinfo{volume}{3}}, \bibinfo{pages}{10} (\bibinfo{year}{2018}).
\newblock \urlprefix\url{https://www.nature.com/articles/s41535-018-0082-7}.

\bibitem{PRLLuis2011}
\bibinfo{author}{Luis, F.} \emph{et~al.}
\newblock \bibinfo{title}{Molecular prototypes for spin-based {CNOT} and {SWAP}
  quantum gates.}
\newblock \emph{\bibinfo{journal}{Phys. Rev. Lett.}}
  \textbf{\bibinfo{volume}{107}}, \bibinfo{pages}{117203}
  (\bibinfo{year}{2011}).
\newblock \urlprefix\url{https://pubmed.ncbi.nlm.nih.gov/22026699/}.

\bibitem{PRLWedge}
\bibinfo{author}{Wedge, C.~J.} \emph{et~al.}
\newblock \bibinfo{title}{Chemical engineering of molecular qubits.}
\newblock \emph{\bibinfo{journal}{Phys. Rev. Lett.}}
  \textbf{\bibinfo{volume}{108}}, \bibinfo{pages}{107204}
  (\bibinfo{year}{2012}).
\newblock
  \urlprefix\url{https://journals.aps.org/prl/abstract/10.1103/PhysRevLett.108.107204}.

\bibitem{Aromi2012}
\bibinfo{author}{Arom\'{\i}, G.}, \bibinfo{author}{Aguil\`{a}, D.},
  \bibinfo{author}{Luis, F.}, \bibinfo{author}{Hill, S.} \&
  \bibinfo{author}{Coronado, E.}
\newblock \bibinfo{title}{Design of magnetic coordination complexes for quantum
  computing.}
\newblock \emph{\bibinfo{journal}{Chem. Soc. Rev.}}
  \textbf{\bibinfo{volume}{41}}, \bibinfo{pages}{537--546}
  (\bibinfo{year}{2012}).
\newblock
  \urlprefix\url{https://pubs.rsc.org/en/content/articlelanding/2012/cs/c1cs15115k/unauth#!divAbstract}.

\bibitem{Aguila2014}
\bibinfo{author}{Aguil\`{a}, D.} \emph{et~al.}
\newblock \bibinfo{title}{Heterodimetallic [{LnLn}'] lanthanide complexes:
  Toward a chemical design of two-qubit molecular spin quantum gates.}
\newblock \emph{\bibinfo{journal}{J. Am. Chem. Soc.}}
  \textbf{\bibinfo{volume}{136}}, \bibinfo{pages}{14215}
  (\bibinfo{year}{2014}).
\newblock \urlprefix\url{https://pubmed.ncbi.nlm.nih.gov/25203521/}.

\bibitem{SciRepNi}
\bibinfo{author}{Chiesa, A.} \emph{et~al.}
\newblock \bibinfo{title}{Molecular nanomagnets with switchable coupling for
  quantum simulation.}
\newblock \emph{\bibinfo{journal}{Sci. Rep.}} \textbf{\bibinfo{volume}{4}},
  \bibinfo{pages}{7423} (\bibinfo{year}{2014}).
\newblock \urlprefix\url{https://www.nature.com/articles/srep07423}.

\bibitem{Ardavan2015}
\bibinfo{author}{Ardavan, A.} \emph{et~al.}
\newblock \bibinfo{title}{Engineering coherent interactions in molecular
  nanomagnet dimers.}
\newblock \emph{\bibinfo{journal}{npj Quantum Inf.}}
  \textbf{\bibinfo{volume}{1}}, \bibinfo{pages}{15012} (\bibinfo{year}{2015}).
\newblock \urlprefix\url{https://www.nature.com/articles/npjqi201512}.

\bibitem{modules}
\bibinfo{author}{Ferrando-Soria, J.} \emph{et~al.}
\newblock \bibinfo{title}{A modular design of molecular qubits to implement
  universal quantum gates.}
\newblock \emph{\bibinfo{journal}{Nat. Commun.}} \textbf{\bibinfo{volume}{7}},
  \bibinfo{pages}{11377} (\bibinfo{year}{2016}).
\newblock \urlprefix\url{https://www.nature.com/articles/ncomms11377}.

\bibitem{Chem}
\bibinfo{author}{Ferrando-Soria, J.} \emph{et~al.}
\newblock \bibinfo{title}{Switchable interaction in molecular double qubits}.
\newblock \emph{\bibinfo{journal}{Chem}} \textbf{\bibinfo{volume}{1}},
  \bibinfo{pages}{727--752} (\bibinfo{year}{2016}).
\newblock
  \urlprefix\url{https://www.sciencedirect.com/science/article/pii/S2451929416301668}.

\bibitem{SIMqubit}
\bibinfo{author}{Ding, Y.-S.}, \bibinfo{author}{Deng, Y.-F.} \&
  \bibinfo{author}{Zheng, Y.-Z.}
\newblock \bibinfo{title}{The rise of single-ion magnets as spin qubits.}
\newblock \emph{\bibinfo{journal}{Magnetochemistry}}
  \textbf{\bibinfo{volume}{2}}, \bibinfo{pages}{40} (\bibinfo{year}{2016}).
\newblock \urlprefix\url{https://www.mdpi.com/2312-7481/2/4/40}.

\bibitem{VO2}
\bibinfo{author}{Atzori, M.} \emph{et~al.}
\newblock \bibinfo{title}{A two-qubit molecular architecture for
  electronmediated nuclear quantum simulation.}
\newblock \emph{\bibinfo{journal}{Chem. Sci.}} \textbf{\bibinfo{volume}{9}},
  \bibinfo{pages}{6183} (\bibinfo{year}{2018}).
\newblock
  \urlprefix\url{https://pubs.rsc.org/en/content/articlelanding/2018/sc/c8sc01695j#!divAbstract}.

\bibitem{Sessoli2019}
\bibinfo{author}{Atzori, M.} \& \bibinfo{author}{Sessoli, R.}
\newblock \bibinfo{title}{The second quantum revolution: Role and challenges of
  molecular chemistry}.
\newblock \emph{\bibinfo{journal}{J. Am. Chem. Soc.}}
  \textbf{\bibinfo{volume}{141}}, \bibinfo{pages}{11339}
  (\bibinfo{year}{2019}).
\newblock \urlprefix\url{https://pubs.acs.org/doi/10.1021/jacs.9b00984}.

\bibitem{Gaitarev}
\bibinfo{author}{Gaita-Ari$\tilde{\text n}$o, A.}, \bibinfo{author}{Luis, F.},
  \bibinfo{author}{Hill, S.} \& \bibinfo{author}{Coronado, E.}
\newblock \bibinfo{title}{Molecular spins for quantum computation.}
\newblock \emph{\bibinfo{journal}{Nat. Chem.}} \textbf{\bibinfo{volume}{11}},
  \bibinfo{pages}{301--309} (\bibinfo{year}{2019}).
\newblock \urlprefix\url{https://www.nature.com/articles/s41557-019-0232-y}.

\bibitem{ErCeEr}
\bibinfo{author}{Macaluso, E.} \emph{et~al.}
\newblock \bibinfo{title}{A heterometallic [{LnLn'Ln}] lanthanide complex as a
  qubit with embedded quantum error correction}.
\newblock \emph{\bibinfo{journal}{Chem. Sci.}} \textbf{\bibinfo{volume}{11}},
  \bibinfo{pages}{10337} (\bibinfo{year}{2020}).
\newblock
  \urlprefix\url{https://pubs.rsc.org/en/content/articlelanding/2020/sc/d0sc03107k#!divAbstract}.

\bibitem{Troiani2008}
\bibinfo{author}{Troiani, F.}, \bibinfo{author}{Bellini, V.} \&
  \bibinfo{author}{Affronte, M.}
\newblock \bibinfo{title}{Decoherence induced by hyperfine interactions with
  nuclear spins in antiferromagnetic molecular rings}.
\newblock \emph{\bibinfo{journal}{Phys. Rev. B}} \textbf{\bibinfo{volume}{77}},
  \bibinfo{pages}{054428} (\bibinfo{year}{2008}).
\newblock \urlprefix\url{https://link.aps.org/doi/10.1103/PhysRevB.77.054428}.

\bibitem{Ghirri2015}
\bibinfo{author}{Ghirri, A.} \emph{et~al.}
\newblock \bibinfo{title}{Coherent spin dynamics in molecular
  $\mathrm{Cr}_8${Z}n wheels}.
\newblock \emph{\bibinfo{journal}{J. Phys. Chem. Lett.}}
  \textbf{\bibinfo{volume}{6}}, \bibinfo{pages}{5062--5066}
  (\bibinfo{year}{2015}).
\newblock \urlprefix\url{https://doi.org/10.1021/acs.jpclett.5b02527}.

\bibitem{Chen2020}
\bibinfo{author}{Chen, J.}, \bibinfo{author}{Hu, C.}, \bibinfo{author}{Stanton,
  J.~F.}, \bibinfo{author}{Hill, S.}, \bibinfo{author}{Cheng, H.-P.} \&
  \bibinfo{author}{Zhang, X.-G.}
\newblock \bibinfo{title}{Decoherence in molecular electron spin qubits:
  Insights from quantum many-body simulations}.
\newblock \emph{\bibinfo{journal}{J. Phys. Chem. Lett.}}
  \textbf{\bibinfo{volume}{11}}, \bibinfo{pages}{2074--2078}
  (\bibinfo{year}{2020}).
\newblock \urlprefix\url{https://doi.org/10.1021/acs.jpclett.0c00193}.

\bibitem{Coish2008}
\bibinfo{author}{Coish, W.~A.}, \bibinfo{author}{Fischer, J.} \&
  \bibinfo{author}{Loss, D.}
\newblock \bibinfo{title}{Exponential decay in a spin bath}.
\newblock \emph{\bibinfo{journal}{Phys. Rev. B}} \textbf{\bibinfo{volume}{77}},
  \bibinfo{pages}{125329} (\bibinfo{year}{2008}).
\newblock \urlprefix\url{https://link.aps.org/doi/10.1103/PhysRevB.77.125329}.

\bibitem{Klauder1962}
\bibinfo{author}{Klauder, J.~R.} \& \bibinfo{author}{Anderson, P.~W.}
\newblock \bibinfo{title}{Spectral diffusion decay in spin resonance
  experiments}.
\newblock \emph{\bibinfo{journal}{Phys. Rev.}} \textbf{\bibinfo{volume}{125}},
  \bibinfo{pages}{912--932} (\bibinfo{year}{1962}).
\newblock
  \urlprefix\url{https://journals.aps.org/pr/abstract/10.1103/PhysRev.125.912}.

\bibitem{Abe2004}
\bibinfo{author}{Abe, E.}, \bibinfo{author}{Itoh, K.~M.},
  \bibinfo{author}{Isoya, J.} \& \bibinfo{author}{Yamasaki, S.}
\newblock \bibinfo{title}{Electron-spin phase relaxation of phosphorus donors
  in nuclear-spin-enriched silicon}.
\newblock \emph{\bibinfo{journal}{Phys. Rev. B}} \textbf{\bibinfo{volume}{70}},
  \bibinfo{pages}{033204} (\bibinfo{year}{2004}).
\newblock \urlprefix\url{https://link.aps.org/doi/10.1103/PhysRevB.70.033204}.

\bibitem{Witzel2005}
\bibinfo{author}{Witzel, W.~M.}, \bibinfo{author}{de~Sousa, R.} \&
  \bibinfo{author}{Das~Sarma, S.}
\newblock \bibinfo{title}{Quantum theory of spectral-diffusion-induced electron
  spin decoherence}.
\newblock \emph{\bibinfo{journal}{Phys. Rev. B}} \textbf{\bibinfo{volume}{72}},
  \bibinfo{pages}{161306} (\bibinfo{year}{2005}).
\newblock \urlprefix\url{https://link.aps.org/doi/10.1103/PhysRevB.72.161306}.

\bibitem{Ardavan2007}
\bibinfo{author}{Ardavan, A.} \emph{et~al.}
\newblock \bibinfo{title}{Will spin-relaxation times in molecular magnets
  permit quantum information processing?}
\newblock \emph{\bibinfo{journal}{Phys. Rev. Lett.}}
  \textbf{\bibinfo{volume}{98}}, \bibinfo{pages}{057201}
  (\bibinfo{year}{2007}).
\newblock
  \urlprefix\url{https://link.aps.org/doi/10.1103/PhysRevLett.98.057201}.

\bibitem{Graham2017}
\bibinfo{author}{Graham, M.~J.}, \bibinfo{author}{Yu, C.-J.},
  \bibinfo{author}{Krzyaniak, M.~D.}, \bibinfo{author}{Wasielewski, M.~R.} \&
  \bibinfo{author}{Freedman, D.~E.}
\newblock \bibinfo{title}{Synthetic approach to determine the effect of nuclear
  spin distance on electronic spin decoherence}.
\newblock \emph{\bibinfo{journal}{J. Am. Chem. Soc.}}
  \textbf{\bibinfo{volume}{139}}, \bibinfo{pages}{3196--3201}
  (\bibinfo{year}{2017}).
\newblock \urlprefix\url{https://doi.org/10.1021/jacs.6b13030}.

\bibitem{Yang2008}
\bibinfo{author}{Yang, W.} \& \bibinfo{author}{Liu, R.-B.}
\newblock \bibinfo{title}{Quantum many-body theory of qubit decoherence in a
  finite-size spin bath}.
\newblock \emph{\bibinfo{journal}{Phys. Rev. B}} \textbf{\bibinfo{volume}{78}},
  \bibinfo{pages}{085315} (\bibinfo{year}{2008}).
\newblock \urlprefix\url{https://link.aps.org/doi/10.1103/PhysRevB.78.085315}.

\bibitem{Chiesa2017}
\bibinfo{author}{Chiesa, A.} \emph{et~al.}
\newblock \bibinfo{title}{Magnetic exchange interactions in the molecular
  nanomagnet $\mathrm{Mn}_{12}$}.
\newblock \emph{\bibinfo{journal}{Phys. Rev. Lett.}}
  \textbf{\bibinfo{volume}{119}}, \bibinfo{pages}{217202}
  (\bibinfo{year}{2017}).
\newblock
  \urlprefix\url{https://link.aps.org/doi/10.1103/PhysRevLett.119.217202}.

\bibitem{Wurger}
\bibinfo{author}{Würger, A.}
\newblock \bibinfo{title}{Magnetic relaxation of mesoscopic molecules}.
\newblock \emph{\bibinfo{journal}{J. Phys.: Condens. Matterr}}
  \textbf{\bibinfo{volume}{10}}, \bibinfo{pages}{10075--10099}
  (\bibinfo{year}{1998}).
\newblock \urlprefix\url{https://doi.org/10.1088/0953-8984/10/44/014}.

\bibitem{BaderChemComm}
\bibinfo{author}{Bader, K.}, \bibinfo{author}{Winkler, M.} \&
  \bibinfo{author}{van Slageren, J.}
\newblock \bibinfo{title}{Tuning of molecular qubits: very long coherence and
  spin–lattice relaxation times}.
\newblock \emph{\bibinfo{journal}{Chem. Commun.}}
  \textbf{\bibinfo{volume}{52}}, \bibinfo{pages}{3623--3626}
  (\bibinfo{year}{2016}).
\newblock \urlprefix\url{http://dx.doi.org/10.1039/C6CC00300A}.

\bibitem{Takahashi}
\bibinfo{author}{Takahashi, S.}, \bibinfo{author}{van Tol, J.},
  \bibinfo{author}{Beedle, C.~C.}, \bibinfo{author}{Hendrickson, D.~N.},
  \bibinfo{author}{Brunel, L.-C.} \& \bibinfo{author}{Sherwin, M.~S.}
\newblock \bibinfo{title}{Coherent manipulation and decoherence of {$S$}=10
  single-molecule magnets}.
\newblock \emph{\bibinfo{journal}{Phys. Rev. Lett.}}
  \textbf{\bibinfo{volume}{102}}, \bibinfo{pages}{087603}
  (\bibinfo{year}{2009}).
\newblock
  \urlprefix\url{https://link.aps.org/doi/10.1103/PhysRevLett.102.087603}.

\bibitem{Stamp}
\bibinfo{author}{Stamp, P. C.~E.} \& \bibinfo{author}{Tupitsyn, I.~S.}
\newblock \bibinfo{title}{Coherence window in the dynamics of quantum
  nanomagnets}.
\newblock \emph{\bibinfo{journal}{Phys. Rev. B}} \textbf{\bibinfo{volume}{69}},
  \bibinfo{pages}{014401} (\bibinfo{year}{2004}).
\newblock \urlprefix\url{https://link.aps.org/doi/10.1103/PhysRevB.69.014401}.

\bibitem{Yao2006}
\bibinfo{author}{Yao, W.}, \bibinfo{author}{Liu, R.-B.} \&
  \bibinfo{author}{Sham, L.~J.}
\newblock \bibinfo{title}{Theory of electron spin decoherence by interacting
  nuclear spins in a quantum dot}.
\newblock \emph{\bibinfo{journal}{Phys. Rev. B}} \textbf{\bibinfo{volume}{74}},
  \bibinfo{pages}{195301} (\bibinfo{year}{2006}).
\newblock \urlprefix\url{https://link.aps.org/doi/10.1103/PhysRevB.74.195301}.

\bibitem{Yang2009}
\bibinfo{author}{Yang, W.} \& \bibinfo{author}{Liu, R.-B.}
\newblock \bibinfo{title}{Quantum many-body theory of qubit decoherence in a
  finite-size spin bath. {II}. ensemble dynamics}.
\newblock \emph{\bibinfo{journal}{Phys. Rev. B}} \textbf{\bibinfo{volume}{79}},
  \bibinfo{pages}{115320} (\bibinfo{year}{2009}).
\newblock \urlprefix\url{https://link.aps.org/doi/10.1103/PhysRevB.79.115320}.

\bibitem{Knill1997}
\bibinfo{author}{Knill, E.} \& \bibinfo{author}{Laflamme, R.}
\newblock \bibinfo{title}{Theory of quantum error-correcting codes}.
\newblock \emph{\bibinfo{journal}{Phys. Rev. A}} \textbf{\bibinfo{volume}{55}},
  \bibinfo{pages}{900--911} (\bibinfo{year}{1997}).
\newblock \urlprefix\url{https://link.aps.org/doi/10.1103/PhysRevA.55.900}.

\bibitem{Gimeno2020}
\bibinfo{author}{Gimeno, I.} \emph{et~al.}
\newblock \bibinfo{title}{Enhanced molecular spin-photon coupling at
  superconducting nanoconstrictions}.
\newblock \emph{\bibinfo{journal}{ACS Nano}} \textbf{\bibinfo{volume}{14}},
  \bibinfo{pages}{8707--8715} (\bibinfo{year}{2020}).
\newblock \urlprefix\url{https://doi.org/10.1021/acsnano.0c03167}.
\newblock \bibinfo{note}{PMID: 32441922}.

\bibitem{Blais2004}
\bibinfo{author}{Blais, A.}, \bibinfo{author}{Huang, R.-S.},
  \bibinfo{author}{Wallraff, A.}, \bibinfo{author}{Girvin, S.~M.} \&
  \bibinfo{author}{Schoelkopf, R.~J.}
\newblock \bibinfo{title}{Cavity quantum electrodynamics for superconducting
  electrical circuits: An architecture for quantum computation}.
\newblock \emph{\bibinfo{journal}{Phys. Rev. A}} \textbf{\bibinfo{volume}{69}},
  \bibinfo{pages}{062320} (\bibinfo{year}{2004}).
\newblock \urlprefix\url{https://link.aps.org/doi/10.1103/PhysRevA.69.062320}.

\bibitem{Blais2021}
\bibinfo{author}{Blais, A.}, \bibinfo{author}{Grimsmo, A.~L.},
  \bibinfo{author}{Girvin, S.~M.} \& \bibinfo{author}{Wallraff, A.}
\newblock \bibinfo{title}{Circuit quantum electrodynamics}.
\newblock \emph{\bibinfo{journal}{Rev. Mod. Phys.}}
  \textbf{\bibinfo{volume}{93}}, \bibinfo{pages}{025005}
  (\bibinfo{year}{2021}).
\newblock
  \urlprefix\url{https://link.aps.org/doi/10.1103/RevModPhys.93.025005}.

\bibitem{Krantz2019}
\bibinfo{author}{Krantz, P.}, \bibinfo{author}{Kjaergaard, M.},
  \bibinfo{author}{Yan, F.}, \bibinfo{author}{Orlando, T.~P.},
  \bibinfo{author}{Gustavsson, S.} \& \bibinfo{author}{Oliver, W.~D.}
\newblock \bibinfo{title}{A quantum engineer's guide to superconducting
  qubits}.
\newblock \emph{\bibinfo{journal}{Appl. Phys. Rev.}}
  \textbf{\bibinfo{volume}{6}}, \bibinfo{pages}{021318} (\bibinfo{year}{2019}).
\newblock \urlprefix\url{https://doi.org/10.1063/1.5089550}.

\bibitem{Jenkins2016}
\bibinfo{author}{Jenkins, M.~D.}, \bibinfo{author}{Zueco, D.},
  \bibinfo{author}{Roubeau, O.}, \bibinfo{author}{Aromí, G.},
  \bibinfo{author}{Majer, J.} \& \bibinfo{author}{Luis, F.}
\newblock \bibinfo{title}{A scalable architecture for quantum computation with
  molecular nanomagnets}.
\newblock \emph{\bibinfo{journal}{Dalton Trans.}}
  \textbf{\bibinfo{volume}{45}}, \bibinfo{pages}{16682--16693}
  (\bibinfo{year}{2016}).
\newblock \urlprefix\url{http://dx.doi.org/10.1039/C6DT02664H}.

\bibitem{Carretta2021}
\bibinfo{author}{Carretta, S.}, \bibinfo{author}{Zueco, D.},
  \bibinfo{author}{Chiesa, A.}, \bibinfo{author}{Gomez-Leon, A.} \&
  \bibinfo{author}{Luis, F.}
\newblock \bibinfo{title}{A perspective on scaling up quantum computation with
  molecular spins}.
\newblock \emph{\bibinfo{journal}{Appl. Phys. Lett.}}
  \textbf{\bibinfo{volume}{118}}, \bibinfo{pages}{240501}
  (\bibinfo{year}{2021}).
\newblock \urlprefix\url{https://aip.scitation.org/doi/10.1063/5.0053378}.

\bibitem{Royer2018}
\bibinfo{author}{Royer, B.}, \bibinfo{author}{Puri, S.} \&
  \bibinfo{author}{Blais, A.}
\newblock \bibinfo{title}{Qubit parity measurement by parametric driving in
  circuit {QED}}.
\newblock \emph{\bibinfo{journal}{Sci. Adv.}} \textbf{\bibinfo{volume}{4}},
  \bibinfo{pages}{eaau1695} (\bibinfo{year}{2018}).
\newblock
  \urlprefix\url{https://advances.sciencemag.org/content/4/11/eaau1695}.

\bibitem{Motzoi2009}
\bibinfo{author}{Motzoi, F.}, \bibinfo{author}{Gambetta, J.~M.},
  \bibinfo{author}{Rebentrost, P.} \& \bibinfo{author}{Wilhelm, F.~K.}
\newblock \bibinfo{title}{Simple pulses for elimination of leakage in weakly
  nonlinear qubits}.
\newblock \emph{\bibinfo{journal}{Phys. Rev. Lett.}}
  \textbf{\bibinfo{volume}{103}}, \bibinfo{pages}{110501}
  (\bibinfo{year}{2009}).
\newblock
  \urlprefix\url{https://link.aps.org/doi/10.1103/PhysRevLett.103.110501}.

\bibitem{Theis2018}
\bibinfo{author}{Theis, L.~S.}, \bibinfo{author}{Motzoi, F.},
  \bibinfo{author}{Machnes, S.} \& \bibinfo{author}{Wilhelm, F.~K.}
\newblock \bibinfo{title}{Counteracting systems of diabaticities using {DRAG}
  controls: The status after 10 years}.
\newblock \emph{\bibinfo{journal}{Europhys. Lett.}}
  \textbf{\bibinfo{volume}{123}}, \bibinfo{pages}{60001}
  (\bibinfo{year}{2018}).
\newblock \urlprefix\url{https://doi.org/10.1209/0295-5075/123/60001}.

\bibitem{Werninghaus2021}
\bibinfo{author}{Werninghaus, M.}, \bibinfo{author}{Egger, D.~J.},
  \bibinfo{author}{Roy, F.}, \bibinfo{author}{Machnes, S.},
  \bibinfo{author}{Wilhelm, F.~K.} \& \bibinfo{author}{Filipp, S.}
\newblock \bibinfo{title}{{Leakage reduction in fast superconducting qubit
  gates via optimal control}}.
\newblock \emph{\bibinfo{journal}{npj Quantum Inf.}}
  \textbf{\bibinfo{volume}{7}}, \bibinfo{pages}{14} (\bibinfo{year}{2021}).
\newblock \urlprefix\url{https://doi.org/10.1038/s41534-020-00346-2}.

\bibitem{Walter2017}
\bibinfo{author}{Walter, T.} \emph{et~al.}
\newblock \bibinfo{title}{Rapid high-fidelity single-shot dispersive readout of
  superconducting qubits}.
\newblock \emph{\bibinfo{journal}{Phys. Rev. Appl.}}
  \textbf{\bibinfo{volume}{7}}, \bibinfo{pages}{054020} (\bibinfo{year}{2017}).
\newblock
  \urlprefix\url{https://link.aps.org/doi/10.1103/PhysRevApplied.7.054020}.

\bibitem{vanSlageren2006}
\bibinfo{author}{van Slageren, J.} \emph{et~al.}
\newblock \bibinfo{title}{Static and dynamic magnetic properties of an
  {$[{\mathrm{Fe}}_{13}]$} cluster}.
\newblock \emph{\bibinfo{journal}{Phys. Rev. B}} \textbf{\bibinfo{volume}{73}},
  \bibinfo{pages}{014422} (\bibinfo{year}{2006}).
\newblock \urlprefix\url{https://link.aps.org/doi/10.1103/PhysRevB.73.014422}.

\bibitem{Schnack2010}
\bibinfo{author}{Schnack, J.}
\newblock \bibinfo{title}{Effects of frustration on magnetic molecules: a
  survey from {O}livier {K}ahn until today}.
\newblock \emph{\bibinfo{journal}{Dalton Trans.}}
  \textbf{\bibinfo{volume}{39}}, \bibinfo{pages}{4677--4686}
  (\bibinfo{year}{2010}).
\newblock \urlprefix\url{https://doi.org/10.1039/B925358K}.

\bibitem{Adelnia2015}
\bibinfo{author}{Adelnia, F.} \emph{et~al.}
\newblock \bibinfo{title}{{Low temperature magnetic properties and spin
  dynamics in single crystals of $\mathrm{Cr}_8$Zn antiferromagnetic molecular
  rings}}.
\newblock \emph{\bibinfo{journal}{J. Chem. Phys.}}
  \textbf{\bibinfo{volume}{143}}, \bibinfo{pages}{244321}
  (\bibinfo{year}{2015}).
\newblock \urlprefix\url{https://doi.org/10.1063/1.4938086}.

\bibitem{Baker2016}
\bibinfo{author}{Baker, M.~L.} \emph{et~al.}
\newblock \bibinfo{title}{Studies of a large odd-numbered odd-electron metal
  ring: Inelastic neutron scattering and muon spin relaxation spectroscopy of
  {Cr}$_8${Mn}}.
\newblock \emph{\bibinfo{journal}{Chem. Eur. J.}}
  \textbf{\bibinfo{volume}{22}}, \bibinfo{pages}{1779--1788}
  (\bibinfo{year}{2016}).
\newblock
  \urlprefix\url{https://chemistry-europe.onlinelibrary.wiley.com/doi/abs/10.1002/chem.201503431}.

\bibitem{Bravyi2011}
\bibinfo{author}{Bravyi, S.}, \bibinfo{author}{DiVincenzo, D.~P.} \&
  \bibinfo{author}{Loss, D.}
\newblock \bibinfo{title}{Schrieffer–{W}olff transformation for quantum
  many-body systems}.
\newblock \emph{\bibinfo{journal}{Ann. Phys.}} \textbf{\bibinfo{volume}{326}},
  \bibinfo{pages}{2793--2826} (\bibinfo{year}{2011}).
\newblock
  \urlprefix\url{https://www.sciencedirect.com/science/article/pii/S0003491611001059}.

\bibitem{Maze2008}
\bibinfo{author}{Maze, J.~R.}, \bibinfo{author}{Taylor, J.~M.} \&
  \bibinfo{author}{Lukin, M.~D.}
\newblock \bibinfo{title}{Electron spin decoherence of single nitrogen-vacancy
  defects in diamond}.
\newblock \emph{\bibinfo{journal}{Phys. Rev. B}} \textbf{\bibinfo{volume}{78}},
  \bibinfo{pages}{094303} (\bibinfo{year}{2008}).
\newblock \urlprefix\url{https://link.aps.org/doi/10.1103/PhysRevB.78.094303}.

\bibitem{Dalessandro2007}
\bibinfo{author}{D'Alessandro, D.}
\newblock \emph{\bibinfo{title}{Introduction to Quantum Control and Dynamics}}.
\newblock Chapman \& Hall/CRC Applied Mathematics \& Nonlinear Science
  (\bibinfo{publisher}{CRC Press}, \bibinfo{year}{2007}).
\newblock \urlprefix\url{https://books.google.co.uk/books?id=e5M0id5enzQC}.

\end{thebibliography}

\noindent{\bf REFERENCES}\\

\end{document}